\documentclass[aps,amsmath,preprint,epsf,superscriptaddress,nofootinbib]{revtex4-1}
\pdfoutput=1
\usepackage{graphicx}
\usepackage{bm}
\usepackage{color}
\usepackage{amsmath,amssymb,amsfonts}	
\usepackage{hyperref}
\usepackage{setspace}
\usepackage[samesize]{cancel}

\def\slashchar#1{\setbox0=\hbox{$#1$} 
\dimen0=\wd0 
\setbox1=\hbox{/} \dimen1=\wd1 
\ifdim\dimen0>\dimen1 
\rlap{\hbox to \dimen0{\hfil/\hfil}} 
#1 
\else 
\rlap{\hbox to \dimen1{\hfil$#1$\hfil}} 
/ 
\fi}

\begin{document}

\def\a{\alpha}
\def\b{\beta}
\def\c{\varepsilon}
\def\d{\delta}
\def\e{\epsilon}
\def\f{\phi}
\def\g{\gamma}
\def\h{\theta}
\def\k{\kappa}
\def\l{\lambda}
\def\m{\mu}
\def\n{\nu}
\def\p{\psi}
\def\q{\partial}
\def\r{\rho}
\def\s{\sigma}
\def\t{\tau}
\def\u{\upsilon}
\def\v{\varphi}
\def\w{\omega}
\def\x{\xi}
\def\y{\eta}
\def\z{\zeta}
\def\D{\Delta}
\def\G{\Gamma}
\def\H{\Theta}
\def\L{\Lambda}
\def\F{\Phi}
\def\P{\Psi}
\def\S{\Sigma}
\def\hyphen{\mathchar`-}

\def\o{\over}
\def\beq{\begin{eqnarray}}
\def\eeq{\end{eqnarray}}
\newcommand{\gsim}{ \mathop{}_{\textstyle \sim}^{\textstyle >} }
\newcommand{\lsim}{ \mathop{}_{\textstyle \sim}^{\textstyle <} }
\newcommand{\vev}[1]{ \left\langle {#1} \right\rangle }
\newcommand{\bra}[1]{ \langle {#1} | }
\newcommand{\ket}[1]{ | {#1} \rangle }
\newcommand{\EV}{ {\rm eV} }
\newcommand{\KEV}{ {\rm keV} }
\newcommand{\MEV}{ {\rm MeV} }
\newcommand{\GEV}{ {\rm GeV} }
\newcommand{\TEV}{ {\rm TeV} }
\def\diag{\mathop{\rm diag}\nolimits}
\def\Spin{\mathop{\rm Spin}}
\def\SO{\mathop{\rm SO}}
\def\O{\mathop{\rm O}}
\def\SU{\mathop{\rm SU}}
\def\U{\mathop{\rm U}}
\def\Sp{\mathop{\rm Sp}}
\def\SL{\mathop{\rm SL}}
\def\tr{\mathop{\rm tr}}
\def\mpl{M_{PL}}

\def\IJMP{Int.~J.~Mod.~Phys. }
\def\MPL{Mod.~Phys.~Lett. }
\def\NP{Nucl.~Phys. }
\def\PL{Phys.~Lett. }
\def\PR{Phys.~Rev. }
\def\PRL{Phys.~Rev.~Lett. }
\def\PTP{Prog.~Theor.~Phys. }
\def\ZP{Z.~Phys. }


\baselineskip 0.7cm

\preprint{IPMU16-0140}
\bigskip

\title{Revisiting gravitino dark matter in thermal leptogenesis}

\author{Masahiro Ibe}
\email[e-mail: ]{ibe@icrr.u-tokyo.ac.jp}
\affiliation{Kavli IPMU (WPI), UTIAS, The University of Tokyo, Kashiwa, Chiba 277-8583, Japan}
\affiliation{ICRR, The University of Tokyo, Kashiwa, Chiba 277-8582, Japan}
\author{Motoo Suzuki}
\email[e-mail: ]{m0t@icrr.u-tokyo.ac.jp}
\affiliation{Kavli IPMU (WPI), UTIAS, The University of Tokyo, Kashiwa, Chiba 277-8583, Japan}
\affiliation{ICRR, The University of Tokyo, Kashiwa, Chiba 277-8582, Japan}
\author{Tsutomu T. Yanagida}
\email[e-mail: ]{tsutomu.tyanagida@ipmu.jp}
\affiliation{Kavli IPMU (WPI), UTIAS, The University of Tokyo, Kashiwa, Chiba 277-8583, Japan}

\begin{abstract}
In this paper, we revisit the gravitino dark matter scenario in the presence of the bilinear $R$-parity violating interaction.
In particular, we discuss a consistency with the thermal leptogenesis.
For a high reheating temperature required for the thermal leptogenesis,  the gravitino dark matter tends to be overproduced, which puts a severe upper limit on the gluino mass.
As we will show, a large portion of parameter space of the gravitino dark matter scenario has been excluded by 
combining the constraints from the gravitino abundance and the null results of the searches for 
the superparticles at the LHC experiments.
In particular, the models with the stau (and other charged slepton) NLSP has been almost excluded by the searches for the long-lived charged particles at the LHC
unless the required reheating temperature is somewhat lowered by assuming, for example, a  degenerated right-handed neutrino mass spectrum.
\end{abstract}

\maketitle

\section{Introduction} 
For decades, supersymmetry has been widely studied as one of the top candidates for physics beyond the Standard Model (SM)
which allows a vast separation of low energy scales from high energy scales such as the Planck scale.
The precise unification of the gauge coupling constants at the scale of the grand unified theory (GUT) also strongly supports
the minimal supersymmetric standard model (MSSM).
In addition, when the $R$-parity~\cite{Fayet:1976et,*Fayet:1977yc} is imposed  to forbid baryon ($B$) and lepton ($L$) number violating interactions, 
the lightest supersymmetric particle (LSP) becomes a good candidate for dark matter.

Although the $R$-parity is very important phenomenologically,  understanding of its origin remains as an open question~\cite{Hall:1983id} .
In fact, in view of general discussion that all global symmetries are necessarily broken by quantum gravity 
effects~\cite{Hawking:1987mz,Lavrelashvili:1987jg,Giddings:1988cx,Coleman:1988tj,Gilbert:1989nq,Banks:2010zn},
the $R$-parity is potentially violated unless it is embedded in gauged symmetries.%
\footnote{Discrete subgroups of the gauge symmetries are immune to quantum gravitational effects~\cite{Krauss:1988zc,Preskill:1990bm,Preskill:1991kd,Banks:1991xj}. }

A popular framework for such embedding is to identify the $R$-parity 
(or the matter parity~\cite{Dimopoulos:1981zb,Weinberg:1981wj,Sakai:1981pk,Dimopoulos:1981dw}) 
with a discrete ${\mathbb Z}_2$ subgroup of the gauged $U(1)_{B-L}$ symmetry,
which naturally emerges once we introduce right-handed neutrinos required for the seesaw mechanism~\cite{Yanagida:1979as,Ramond:1979py}~\cite[see also][]{Minkowski:1977sc}.
There, the Majorana masses of the right-handed neutrinos are induced when the $U(1)_{B-L}$ symmetry is broken down to its ${\mathbb Z}_2$ subgroup spontaneously.

This framework is, however, known to have a tension with perturbative $SO(10)$ GUT.
There, the Majorana mass terms of the right-handed neutrinos require the vacuum expectation value (VEV)
of fields in ${\bf 126}$ or larger representations of $SO(10)$.
However, an introduction of fields in such large representations causes a rapid blow up of the $SO(10)$ gauge coupling constant just above the GUT scale.
To avoid this problem, it is often assumed that $U(1)_{B-L}$ by a VEV of $\overline{\bf 16}$ representation 
with which the Majorana masses are given by $\vev{\overline{\bf 16}}^2$.
In this case, the ${\mathbb Z}_2$ subgroup of $U(1)_{B-L}$ does not remain unbroken, and hence, no exact $R$-parity remains.%
\footnote{See \cite{Watari:2015ysa} for a recent discussion on  $R$-parity violation in string theory. }
Rather, this argument opens up a new framework%
\footnote{This does not preclude the $R$-parity originating from  symmetries other than $U(1)_{B-L}$ though.}
where small $R$-parity violation effects are tied with $U(1)_{B-L}$ breaking 
as pursued in Refs.~\cite{Buchmuller:2007ui,Schmidt:2010yu}.

Once  $R$-parity violation is accepted, the gravitino LSP has a definite advantage to be a candidate for dark matter. 
Compared with other LSP candidates, the gravitino LSP can have a much longer lifetime even in the presence of $R$-parity violation~\cite{Takayama:2000uz,Moreau:2001sr}.

In this paper, we discuss gravitino dark matter in the presence of the $R$-parity violating interactions.
In particular, we revisit a consistency with the thermal leptogenesis~\cite{Fukugita:1986hr}~\cite[see][for review]{Giudice:2003jh,Buchmuller:2005eh,Davidson:2008bu}.
For a high reheating temperature required for the thermal leptogenesis,  the gravitino dark matter tends to be overproduced, which puts a severe upper limit on the gluino 
mass~\cite{Bolz:2000fu,Buchmuller:2007ui,Hamaguchi:2009sz}.
As we will show, a large portion of parameter space of the gravitino dark matter scenario has been excluded by 
combining the constraints from the gravitino abundance and the null results of the searches for 
the superparticles at the LHC experiments.

The organization of this paper is as follows. 
In Sec.\,\ref{sec:R-parity breaking}, we briefly review  ${R}$-parity violation in the MSSM. 
We also review the gravitino properties in the presence of $R$-parity violation.
In Sec.\,\ref{sec:thermal leptogenesis}, we discuss a consistency between the gravitino dark matter scenario and thermal leptogenesis scenario. 
There, we also discuss the constraints from the LHC experiments.
Final section is devoted to our conclusions and discussions.

\section{R-parity violation and the gravitino dark matter}
\label{sec:R-parity breaking}
Let us briefly review  ${R}$-parity violation in the MSSM (see  \cite{Barbier:2004ez} for a detailed review). 
The general renormalizable ${R}$-parity violating superpotential is given by
\begin{eqnarray}
\label{eq:Rv}
W_{{\cancel{R}}}=\frac{1}{2} \lambda_{ijk} L_L{}_i L_L{}_j \bar{E}_R{}_k + \lambda^{\prime}_{ijk} L_L{}_i Q_L{}_j\bar{D}_R{}_k
+\frac{1}{2}\lambda^{\prime\prime}_{ijk} \bar{U}_R{}_i\bar{D}_R{}_j \bar{D}_R{}_k +\mu^{\prime}_iL_L{}_iH_u\ ,
\end{eqnarray}
where $i,j,k = 1,2,3$ denote the family indices of the matter fields.
The coefficients  $\lambda^{(\prime, \prime\prime)}$ and $\mu^{\prime}$ are dimensionless and dimensionful parameters of $R$-parity violation, respectively.
The third term violates the $B$-number while the other terms violate the $L$-number.

The most universal constraints on  $R$-parity violation come from cosmology.
In the presence of the $B$ and/or $L$-number violating processes induced by $R$-parity violation,
the baryon asymmetry generated before the electroweak phase transition would be washed out.
To avoid this problem, the $R$-parity violating parameters are constrained to be,
\begin{eqnarray}
\label{eq:cosmology}
\lambda, \, \lambda^\prime, \, \lambda^{\prime\prime}, \, \mu^\prime/\mu < {\cal O}(10^{-(6-7)}) \ ,
\end{eqnarray}
where $\mu$ denotes the $R$-parity conserving $\mu$-parameter in the TeV range~\cite{Christodoulakis:1990zy,Fischler:1990gn,Dreiner:1992vm,Endo:2009cv}.%
\footnote{See Ref.~\cite{Higaki:2014eda}, for baryogengesis~\cite{Affleck:1984fy,Dine:1995kz} in the presence of the $R$-parity violation. } 
Hereafter, we suppress the family indices for simplicity.

The bilinear $R$-parity terms in Eq.\,(\ref{eq:Rv}) are also constrained from the neutrino mass~\cite{Barbier:2004ez}.
By taking the cosmological upper limit on the neutrino mass, $ \sum_im_{\nu_i} \lesssim 0.183$\,eV (at $95$\% CL)~\cite{Giusarma:2016phn},
the constraint is given by,
\begin{eqnarray}
\label{eq:nmass}
\sum_i \frac{\mu^{\prime 2}}{\mu^2} \lesssim 2\times 10^{-11}\tan^2\beta \left( \frac{M_{2}}{1\,\rm TeV}\right)
\left(\frac{M_{1}}{M_{1} c_W^2 + M_{2} s_W^2}\right) \ .
\end{eqnarray}
Here, $M_{1,2}$ are the soft supersymmetry breaking masses of the bino and the wino, respectively, 
$\tan\b$ is the ratio between the VEVs of the two Higgs doublets, and $s_W^2$ denotes the weak mixing angle, $\sin^2\theta_W$, with $c_W^2 = 1-s_W^2$.
It should be noted that we take the basis of the Higgs bosons and the sleptons, $(H_d, \tilde{L}_i)$, so that no sleptons obtain VEVs (see  \cite{Barbier:2004ez} for details).

Now, let us briefly discuss  $R$-parity violation tied to  $U(1)_{B-L}$ breaking.
In particular, we focus on models where  the effects of $R$-parity violation in the MSSM appear 
through tiny VEVs of the right-handed sneutrinos, $\vev{\tilde{N}_R}$~\cite{Schmidt:2010yu,Ibe:2013nka} 
(see also the appendix\,\ref{sec:model}).
In this class of models, the $R$-parity violating parameters are generated as%
\footnote{In the basis of $(H_d, \tilde{L}_i)$ where no sleptons obtain VEVs,
the trilinear terms are also generated as $\lambda \sim  \lambda^\prime \sim \mu'/\mu\ $.
} 
\begin{eqnarray}
\mu^\prime  \sim y_\nu \text{$\vev{\tilde{N}_R}$}\ . \end{eqnarray}
Thus, the constraints in Eqs.\,(\ref{eq:cosmology}) and (\ref{eq:nmass}) on $\lambda^{(\prime)}$ and $\mu^\prime$ can be satisfied as long as $\vev{N_R}$'s
are small.

As an advantageous feature of this class of models, the $B$-violating term, $\lambda^{\prime\prime}$, 
can be further suppressed by additional symmetries (see the appendix\,\ref{sec:model}).
Thus, this class of models can evade the sever constraints  from the null observation of proton decay~\cite{Barbier:2004ez} ,
\begin{eqnarray}
|\lambda^\prime\lambda^{\prime\prime}| \lesssim 10^{-25} \left(\frac{m_{\rm SUSY}}{1\,\rm TeV}\right)^2\ ,
\end{eqnarray}
where $m_{\rm SUSY}$ denotes a typical mass of superparticles.

When the $R$-parity violation effects are dominated by the bilinear terms, 
the gravitino mainly decays into a pair of a $Z$ boson and  a neutrino, a pair of a Higgs boson and a neutrino, 
and a pair of a $W$ boson and a charged lepton. 
The relative branching ratios into those modes converge to $1:1:2$ in the limit of $m_{3/2}\gg m_{Z,W,h}$~\cite{Ishiwata:2008cu,*Ishiwata:2008cv,*Ishiwata:2009vx,Delahaye:2013yqa}.
The decay widths of those modes are roughly given by,
\begin{eqnarray}
2\Gamma[\psi_{3/2} \to Z\nu] \sim 2\Gamma[\psi_{3/2} \to h\nu] 
\sim \Gamma[\psi_{3/2} \to W\ell] \sim  \frac{m_{3/2}^3}{192\pi M_{\rm PL}^2} \left(\frac{\mu^\prime}{\mu}\right)^2\ , 
\end{eqnarray}
leading to the lifetime of the gravitino,
\begin{eqnarray}
\label{eq:tau32}
\tau_{3/2} \simeq  10^{20}\,{\rm sec} \times \left(\frac{1\,\rm TeV}{m_{3/2}}\right)^3 \left(\frac{10^{-7}\mu}{\mu^\prime}\right)^2\ . 
\end{eqnarray}
Here, $m_{3/2}$ denotes the gravitino mass, and $M_{\rm PL} \simeq 2.4\times 10^{18}$\,GeV  the reduced Planck scale.
Therefore, the lifetime of the gravitino can be much longer than the age of the universe, ${\cal O}(10^{17})$ sec, 
for $\mu_i^\prime/\mu \ll 10^{-7}$ for the graivitino in the hundreds GeV to a TeV range.

The gravitino lifetime in the range of Eq.\,(\ref{eq:tau32}) is, however, severely constrained from the observation of
the extragalactic gamma-ray background (EGRB)~\cite{Ibarra:2007wg,Ishiwata:2009dk,Carquin:2015uma,Ando:2015qda,*Ando:2016ang}.%
\footnote{The observations of the neutrino fluxes also constraints the lifetime of the gravitino, which is less stringent than those from the EGRB~\cite{Covi:2008jy}. }
By using 50-month EGRB observation by Fermi-LAT~\cite{Ackermann:2014usa}, 
the lifetime of the gravitino dark matter decaying into a pair of a $W$ boson and a charged lepton is constrained to be $\tau_{3/2} \gtrsim 10^{28}$\,sec
for $m_{3/2} = {\cal O}(100\,{\rm GeV}\mbox{--}\,1\,{\rm TeV})$. 
In the following, we assume that the bilinear $R$-parity violating parameters satisfy
\begin{eqnarray}
\label{eq:EGRB}
 \frac{\mu'}{\mu} \lesssim 10^{-11}\times \left(\frac{1\,\rm TeV}{m_{3/2}}\right)^{3/2}\ .
\end{eqnarray}

\section{gravitino dark matter in thermal leptogensis scenario}
\label{sec:thermal leptogenesis}
\subsection{Thermal gravitino production}

\begin{figure}[t]
\begin{center}
\begin{minipage}{.4\linewidth}
  \includegraphics[width=\linewidth]{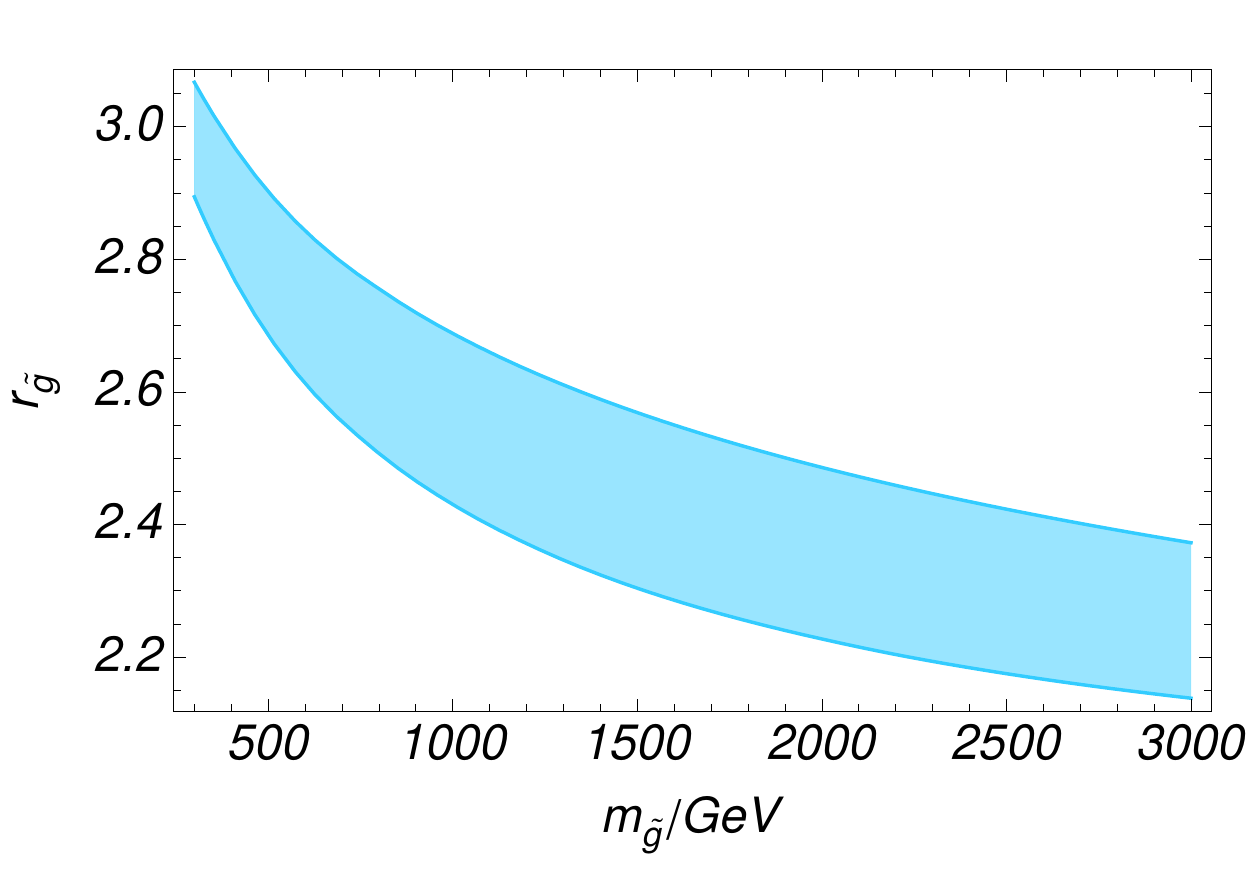}
 \end{minipage}
  \end{center}
\caption{\sl \small
The parameter $r_{\tilde{g}}$ as a function of the physical gluino mass, $m_{\tilde{g}}$. 
The upper and lower bound correspond to $10$ TeV and $1$ TeV squark mass. 
We fix $\beta=10$ although $r_{\tilde g}$ barely depends on $\tan\beta$ nor other model parameters than $m_{1/2}$ and the squark masses.}
\label{fig:rg}
\end{figure}

The productions of the gravitino from the thermal bath are dominated by the QCD process.
The resultant relic density of the gravitino dark matter is given by~\cite{Rychkov:2007uq,Ellis:2015jpg},
\begin{eqnarray}
\label{eq:O32}
\Omega_{3/2}h^2 &\simeq &
0.09 \left(\frac{m_{3/2}}{100\,{\rm GeV}}\right)
\left(\frac{T_R}{10^{10}\,\rm GeV}\right)
\left(
\left(
1 + 0.558 
\frac{r_{\tilde g}^{-2}m_{\tilde g}^2}{m_{3/2}^2}
\right)
\right. 
\nonumber \\
&&\hspace{3cm}
\left.
-0.011\left(
1 + 3.062 
\frac{r_{\tilde g}^{-2}m_{\tilde g}^2}{m_{3/2}^2}
\right)
\log\left[
\frac{T_R}{10^{10}\,\rm GeV}
\right]
\right)
\ ,
\end{eqnarray}
for the universal gaugino mass generated at the GUT scale.%
\footnote{Electroweak contributions to the gravitino production increases the abundance by around $20$\%
when the gaugino masses satisfy the GUT relation~\cite{Pradler:2006qh}.}
Here, the parameter $r_{\tilde g}$ is introduced to translate the universal gaugino mass parameter, $m_{1/2}$, at the GUT scale
to the physical gluino mass, $m_{\tilde g}$,
\begin{eqnarray}
m_{\tilde g} = r_{\tilde g} m_{1/2}\ ,
\end{eqnarray}
which depends on the MSSM parameters. 
In Fig.\ref{fig:rg}, we show $r_{\tilde{g}}$ as a function of the physical gluino mass 
calculated with the code {\tt SOFTSUSY} \cite{Allanach:2001kg} where the upper and lower bound correspond to squark mass $10$ TeV and $1$ TeV.
Here, we fix $\beta=10$ although $r_{\tilde g}$ barely depends on $\tan\beta$ nor other model parameters than $m_{1/2}$ and the squark masses.
In our analysis, we adopt the upper bound of the Fig.\,\ref{fig:rg} for a given $m_{\tilde g}$,
which makes the following analysis conservative.
It should be noted that the relic density in Eq.\,(\ref{eq:O32}) is about a factor two larger than the one in \cite{Bolz:2000fu} 
which is caused by the thermal mass effects of the gluon as discussed in~\cite{Rychkov:2007uq}.

\begin{figure}[t]
\begin{center}
\begin{minipage}{.4\linewidth}
  \includegraphics[width=\linewidth]{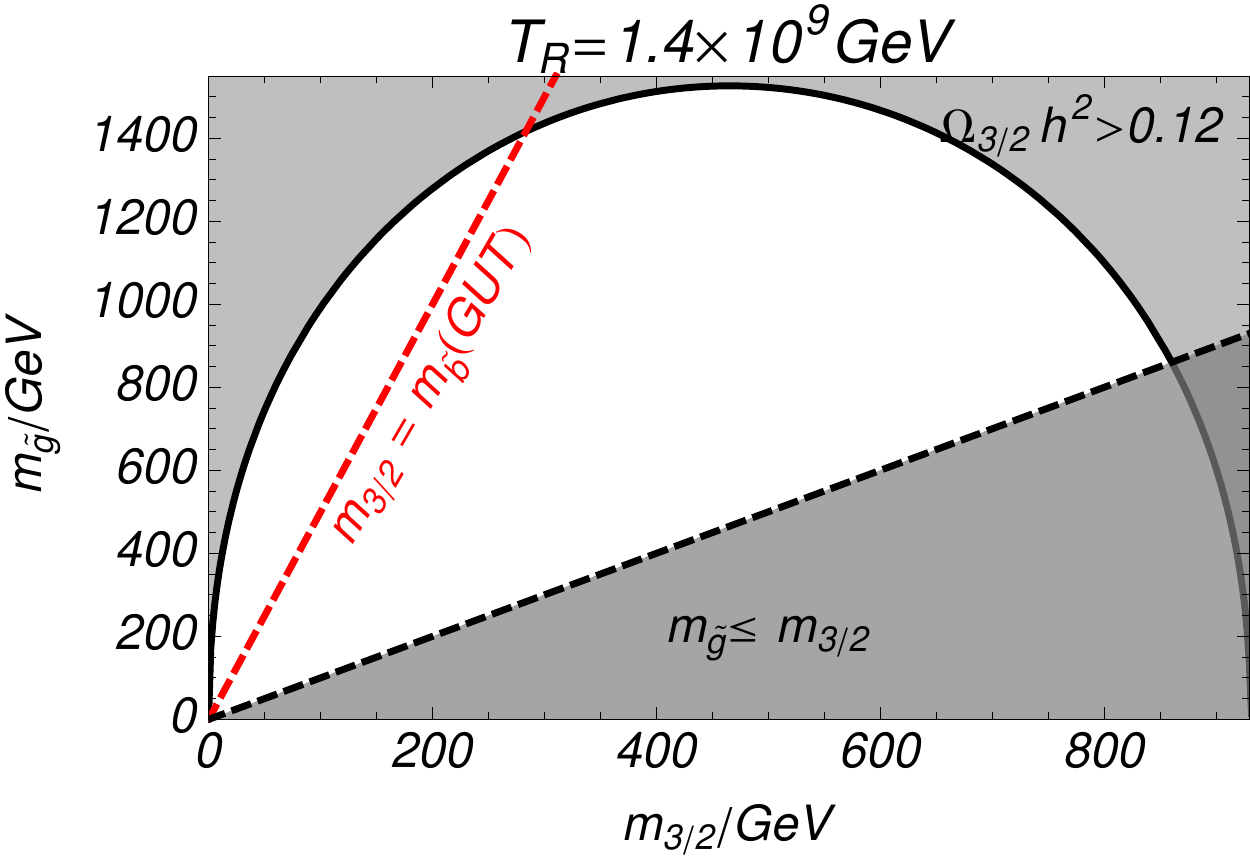}
 \end{minipage}
 \hspace{1cm}
 \begin{minipage}{.4\linewidth}
  \includegraphics[width=\linewidth]{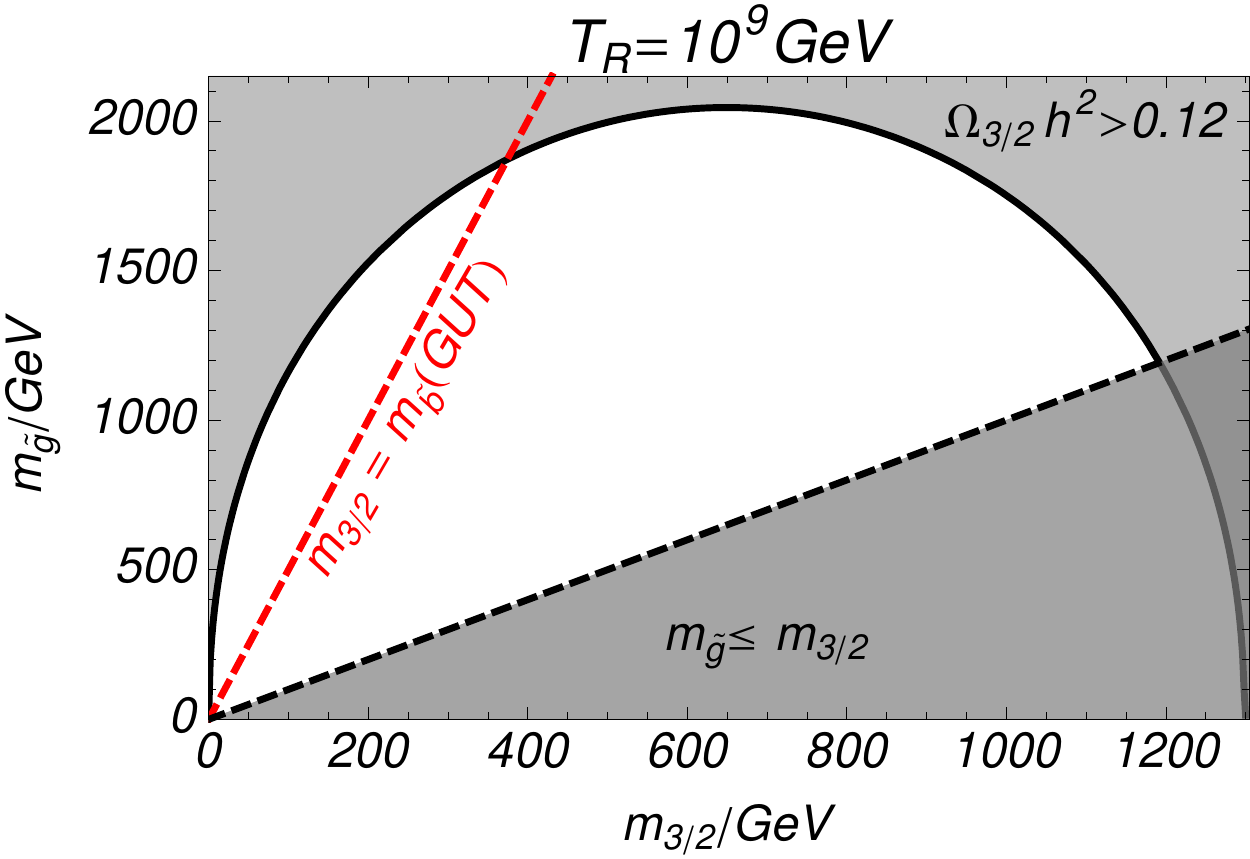}
 \end{minipage}
  \end{center}
\caption{\sl \small
The upper limits on the gluino mass as a function of the gravitino mass, $m_{3/2}$, for given reheating temperatures. 
Below the black dashed lines, the gravitino becomes heavier than the gluino.
The  gray shaded regions are excluded where the gravitino abundance exceeds the observed dark matter abundance.
The gaugino masses can satisfy the GUT relation in the left side of the red dashed line while keeping the gravitino being the LSP.
}
\label{fig:m32_mglui}
\end{figure}

From the relic density in Eq.\,(\ref{eq:O32}), we immediately find that the gluino mass is severely constrained 
from above for successful leptogenesis which requires a high reheating temperature, $T_R \gtrsim 1.4\times 10^{9}$\,GeV~\cite{Antusch:2006gy}.%
\footnote{The definitions of the reheating temperature in \cite{Giudice:2003jh,Antusch:2006gy} 
and in \cite{Ellis:2015jpg} are slightly different and the former is about $30$\% larger for a given
inflaton decay width.}
Here, we assume that the spectrum of the right-handed neutrinos are not degenerated.
In Fig.\,\ref{fig:m32_mglui}, we show the upper limits on the gluino mass for given reheating temperatures.
In the figure, the gray shaded regions  are excluded where the gravitino relic density exceeds
the observed dark matter density, $\Omega h^2 \simeq 0.1198\pm 0.0015$~\cite{Ade:2015lrj}.
The dark matter density can be fully explained on the upper limit on the gluino mass for a given gravitino mass.
The figure shows that the upper limit on the gluino mass is around $1.5$\,TeV for $T_R \gtrsim 1.4\times 10^9$\,GeV.
In the right panel of Fig.\,\ref{fig:m32_mglui}, we also show the upper limit on the gluino mass for $T_R \simeq 10^9$\,GeV
in case that the reheating temperature required for leptogenesis is somewhat relaxed.

Let us comment here that the constraints become much severer when the gaugino masses 
satisfy the GUT relation,
\begin{eqnarray}
m_{\tilde b} : m_{\tilde w} : m_{\tilde g} \sim 1: 2 : 5\,\mbox{--}\,6\ ,
\end{eqnarray}
where $m_{\tilde b, \tilde w}$ are the masses of the bino and the wino, respectively.
When the GUT relation is satisfied, the gravitino mass should be smaller than the bino mass
\begin{eqnarray}
m_{3/2} < m_{\tilde b} \simeq \frac{1}{5} m_{\tilde g}\ ,
\end{eqnarray}
so that the gravitino is the LSP.
In Fig.\,\ref{fig:m32_mglui}, this condition can be satisfied only in the left side of the red dashed line.
There, the limits on the gluino mass are $m_{\tilde g} \lesssim 1.3$\,TeV for $T_R > 1.4\times 10^9$\,GeV and $m_{\tilde g}\lesssim 1.8$\,TeV for $T_R > 10^9$\,GeV,
respectively.

In the right side of the red dashed line, on the other hand, the bino and the wino should be heavier than the GUT relation in order for the gravitino to be the LSP.
Since the heavier bino/wino masses increase the electroweak contributions to the gravitino abundance, the abundance in Eq.\,(\ref{eq:O32})  
underestimates the gravitino abundance in this region.
Therefore, the upper limit on the gluino mass in the right side of the red dashed line is rather conservative.

\subsection{NLSP contributions}
As discussed in \cite{Feng:2003xh,*Feng:2003uy,Fujii:2003nr,Feng:2004zu,*Feng:2004mt}, the late-time decay of the 
next-to-lightest superparticle (NLSP) also contributes to the relic gravitino density,
\begin{eqnarray}
\label{eq:OM2}
\Omega_{3/2}^{\rm tot}h^2 = \Omega_{3/2}h^2 + Br_{3/2}\times \frac{m_{3/2}}{m_{\rm NLSP}} \Omega_{\rm NLSP}h^2\ .
\end{eqnarray}
Here, $\Omega_{\rm NLSP}$ denotes the thermal relic density of the NLSP when it is stable, and $Br_{3/2}$
 the branching fraction of the NLSP into the gravitino.
In the absence of  $R$-parity violation, the NLSP dominantly decays into the gravitino, and hence, $Br_{3/2} =1$.
In this case, the constraints on the gluino mass from the gravitino abundance in Fig.\,\ref{fig:m32_mglui} are severer~\cite{Fujii:2003nr,Heisig:2013sva,Arvey:2015nra}.

In the presence of  $R$-parity violation, on the contrary, $Br_{3/2}$ can be much suppressed
when the effects of $R$-parity violation are sizable.
In fact, the width of the $R$-parity violating decay (via the bilinear $R$-parity violating terms) is roughly given by,
\begin{eqnarray}
\G_{\rm NLSP}^{\cancel R} \simeq \frac{\k}{16\pi} \left(
\frac{\mu^\prime}{\mu}\right)^2m_{{\rm NLSP}}\ ,
\end{eqnarray}
which leads to the lifetime,
\begin{eqnarray}
\tau_{\rm NLSP}^{\cancel R} \simeq 10^{-4}{\rm sec}\times \k^{-1}\left(\frac{10^{-11}\mu}{\mu^\prime}\right)^2\left(\frac{1\, {\rm TeV}}{m_{\rm NLSP}}\right).
\end{eqnarray}
Here, $\k$ represents dependences on the MSSM parameters~\cite{Buchmuller:2007ui,Ishiwata:2008cu}.
This width is much larger than the width of the $R$-parity conserving decay into a pair of the gravitino and the superpartner of the NLSP,
\begin{eqnarray}
\G_{\rm NLSP}^{R} \simeq \frac{1}{48\pi}\frac{m_{\rm{NLSP}}^5}{m_{3/2}^2\mpl^2}\ ,
\end{eqnarray}
which corresponds to
\begin{eqnarray}
\tau_{\rm NLSP}^R \simeq 5\times10^{3}{\rm sec}\left(\frac{m_{3/2}}{100{\rm GeV}}\right)^2\left(\frac{1\,{\rm TeV}}{m_{{\rm NLSP}}}\right)^5\ .
\end{eqnarray}
Therefore, $Br_{3/2}$ is expected to be very small even for small $R$-parity violating bilinear terms as in Eq.\,(\ref{eq:Rrange}).

It should be also noted that the properties of the NLSP are strongly constrained by Big-Bang nucleosynthesis (BBN)~\cite{Kawasaki:2004qu,Jedamzik:2006xz}. 
For the neutralino NLSP, for example, the lifetime should be shorter than $10^{2}$\,sec to avoid 
dissociation of the light elements by the NLSP decays into hadronic showers (especially into nucleons).
For the stau NLSP, on the other hand, the lifetime should be shorter than $10^{3-4}$\,sec to avoid the 
light element dissociation.
By taking those constraints from the BBN into account, we assume that the $R$-parity violating parameters are in the range of,
\begin{eqnarray}
\label{eq:Rrange}
10^{-14}\times
\k^{-1/2}\left(\frac{1\,\rm TeV}{m_{\rm NLSP}}\right)^{1/2}\,
\lesssim
\frac{\mu'}{\mu} \,\lesssim 10^{-11}\times \left(\frac{1\,\rm TeV}{m_{3/2}}\right)^{3/2}\ .
\end{eqnarray}
In the appendix\,\ref{sec:model}, we show a model which leads to the bilinear $R$-parity violation terms in this range.
It should be noted that $Br_{3/2}\ll 1$ in this range of $R$-parity violation, and hence, the NLSP contribution to the gravitino abundance in Eq.\,(\ref{eq:OM2}) is negligible.%
\footnote{For typical size of $\Omega_{\rm NLSP}h^2$, see \cite{Fujii:2003nr}.}

\subsection{Collider Constraints}
\label{sec:colliders}
In this subsection, we discuss the constraints from the superparticle searches at the LHC.
First, let us note that the $R$-parity violating parameters we are interested in are small and we can apply the search strategies for the superparticles at the LHC
in the $R$-conserving case.%
\footnote{See also \cite{Hirsch:2005ag} for the effects of $R$-parity violation on the LHC search for much lighter gravitinos.}
In fact, for the neutralino NLSP, the lifetime is typically given by~\cite{Buchmuller:2007ui,Ishiwata:2008cu},
\begin{eqnarray}
\label{eq:Lneutralino}
c \tau_{\tilde \chi_1^0}\gtrsim10^{6}\,{\rm m}\times\left(\frac{1{\rm TeV}}{m_{\tilde \chi_1^0}}\right)^{3}\left(\frac{10^{-11}\mu}{\mu^\prime}
\right)^{2}
\left(\frac{10}{{\rm tan}\beta}\right)^{2}\ .
\end{eqnarray}
For the stau NLSP, on the other hand, the lifetime is similarly given by~\cite{Buchmuller:2007ui,Ishiwata:2008cu},
\begin{eqnarray}
\label{eq:Lstau}
c\tau_{\tilde \t} \gtrsim 10^{7}\, {\rm m}\times \left(\frac{1{\rm TeV}}{m_{\tilde \t}}\right)\left(\frac{10^{-11}\mu}{\mu^\prime}\right)^{2}
\left(\frac{10}{{\rm tan}\beta}\right)^{2} \ .
\end{eqnarray}
Therefore, the NLSP is stable inside the detectors in both cases. 

\begin{figure}[t]
\begin{center}
\begin{minipage}{.45\linewidth}
  \includegraphics[width=\linewidth]{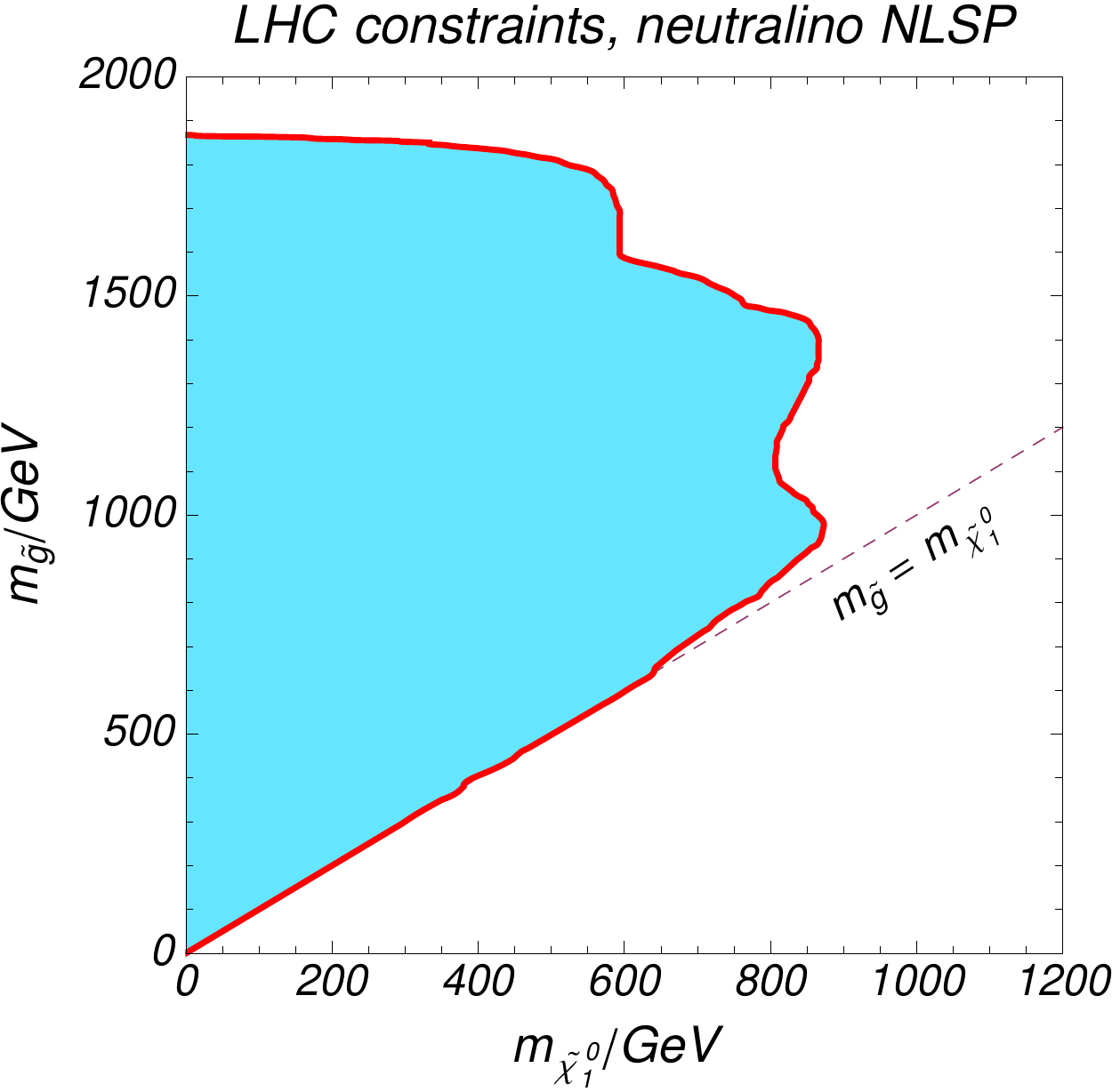}
 \end{minipage}
 \hspace{1cm}
 \begin{minipage}{.45\linewidth}
  \includegraphics[width=\linewidth]{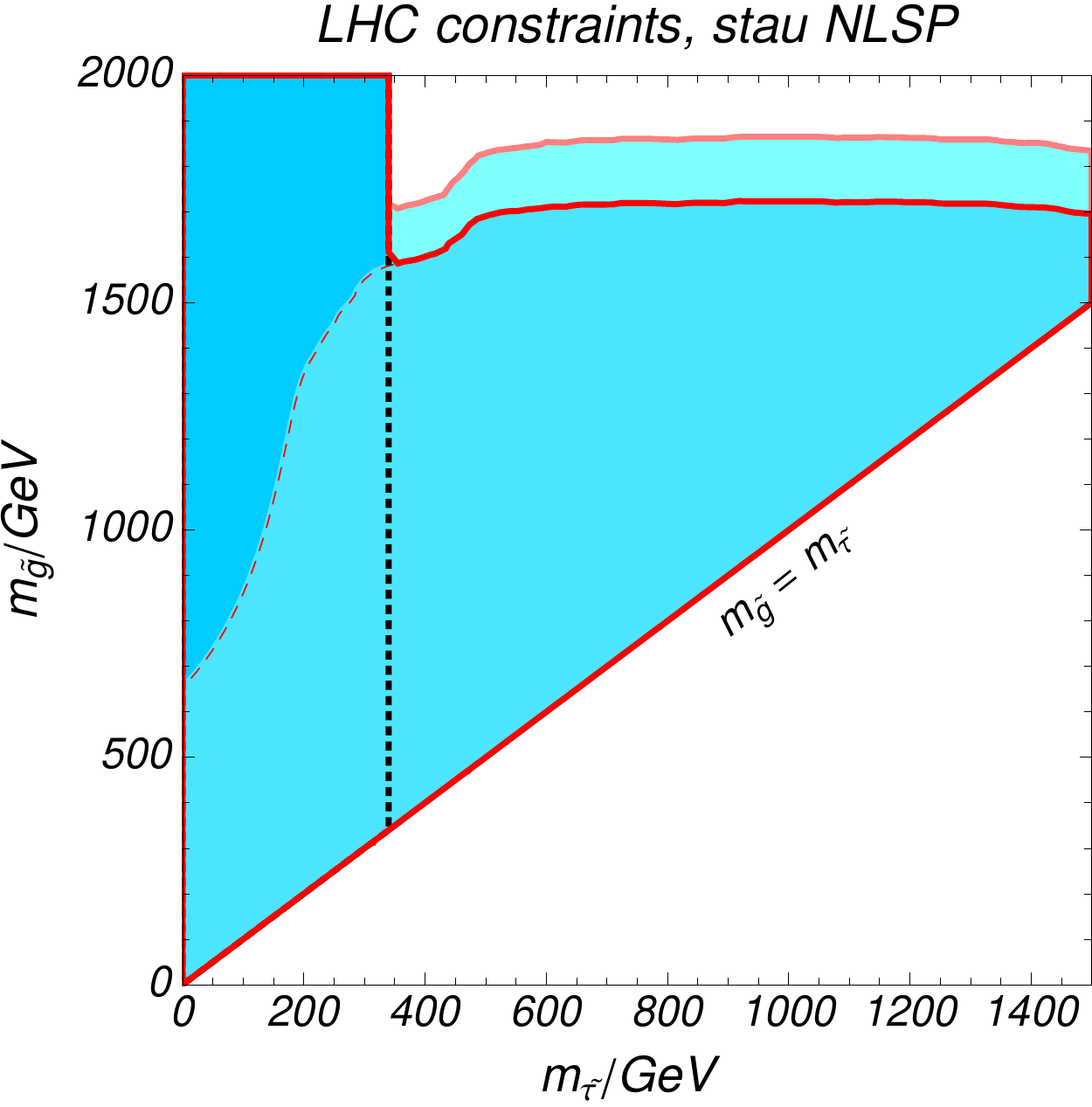}
 \end{minipage}
  \end{center}
\caption{\sl \small Left) The constraints on $(m_{\tilde\chi_1^0}, m_{\tilde g})$  
from the searches for multi-jets with missing momentum extracted from \cite{ATLAS:2016kts}. 
 The neutralino--gluino degenerated region is also excluded 
 by the mono-jet searches up to 600\,GeV~\cite{Arbey:2015hca}.
Right) The constraints on $(m_{\tilde{\t}}, m_{\tilde{g}})$
from the searches for long-lived charged particles.
The constraint on the stau production cross section in \cite{CMS:2015kdx}
is converted to the gluino mass bound by using the gluino NLO+NLL production cross section at 13\,TeV (reduced by 2$\sigma$ theoretical uncertainties)
in \cite{Borschensky:2014cia}.
The region with $m_{\tilde \t} < 340\,$GeV is also excluded by the  long-lived charged particle searches by assuming direct stau production  \cite{CMS:2015kdx}. It is noted that the region below $m_{\tilde{g}}\lesssim 1.5$ TeV with stable gluino where stau and gluino mass is (almost) degenerate is also excluded by R-hadron search \cite{CMS:2015kdx}. In both panels, we assume that the squarks are heavy and decoupled.
In both panels, we assume that constraints are obtained in the limit of decoupled squarks and therefore no dependence of the squark mass is present. 
}
\label{fig:collider}
\end{figure}

Let us begin with the collider constraints in the case of the neutralino NLSP.
Since the neutralino NLSP is stable inside the detectors, we consider the searches for multi-jets with missing momentum.
To derive conservative limits on the gluino mass, we assume that  all the  squarks are 
heavy and decoupled.
In Fig\,\ref{fig:collider}, we show the constraints on the gluino mass and the neutralino mass
at the $95$\% CL which are extracted from the results by the ATLAS collaboration~\cite{ATLAS:2016kts}.%
\footnote{See also \cite{CMS:2016mwj}, for the constraints put by the CMS collaboration.}
Here, the gluino is assumed to decay into two quarks and a neutralino for simplicity.
The figure shows that the lower limit on the gluino mass is around $1.8$\,TeV 
when the neutralino is not degenerated with the gluino.
When the neutralino mass is close to the gluino mass, the constraints become
weaker though the the gluino mass below $1$\,TeV is excluded 
unless the neutralino is highly degenerated with the gluino.
 It should be noted that the neutralino--gluino degenerated region is also excluded 
 by the mono-jet searches up to 600\,GeV~\cite{Arbey:2015hca}.

For the stau NLSP, on the other hand, we consider the long-lived charged particle searches.  
So far, the CMS collaboration puts a lower limit on the mass of the long-lived stau, $m_{\tilde  \t} > 340$\,GeV 
at 95\% CL, by assuming a direct Drell-Yan stau pair production~\cite{CMS:2015kdx}.
The CMS collaboration also puts  upper limits on the production cross section of the stau pairs for 
a given stau mass.
Since the stau production cross section (including the one from the cascade decays of the gluinos) depends on the gluino mass, we can put constraints on 
the gluino mass for a given stau mass.
In Fig.\,\ref{fig:collider}, we show the resultant constraints on $(m_{\tilde \t}, m_{\tilde g})$ plane.
Here, we obtain the constraints by comparing the $95$\% CL limits on the stau production cross section in  \cite{CMS:2015kdx}
with the gluino production cross section in \cite{Borschensky:2014cia}.
The light shaded region denotes the excluded region for the central value of the gluino NLO+NLL production cross section at 13 TeV in \cite{Borschensky:2014cia}, while the darker shaded
region denotes the one for the cross section reduced by $2\sigma$ theoretical uncertainties due to variation of the renormalization and factorization scales and the parton distribution functions. It is noted that the region below $m_{\tilde{g}}\lesssim 1.5$ TeV with stable gluino where stau and gluino mass is (almost) degenerate is also 
 excluded by the $R$-hadron searches which we discuss more detail later in this section.
In the following analysis, we use the later constraint for conservative estimation.

\begin{figure}[t]
\begin{center}
\begin{minipage}{.45\linewidth}
  \includegraphics[width=\linewidth]{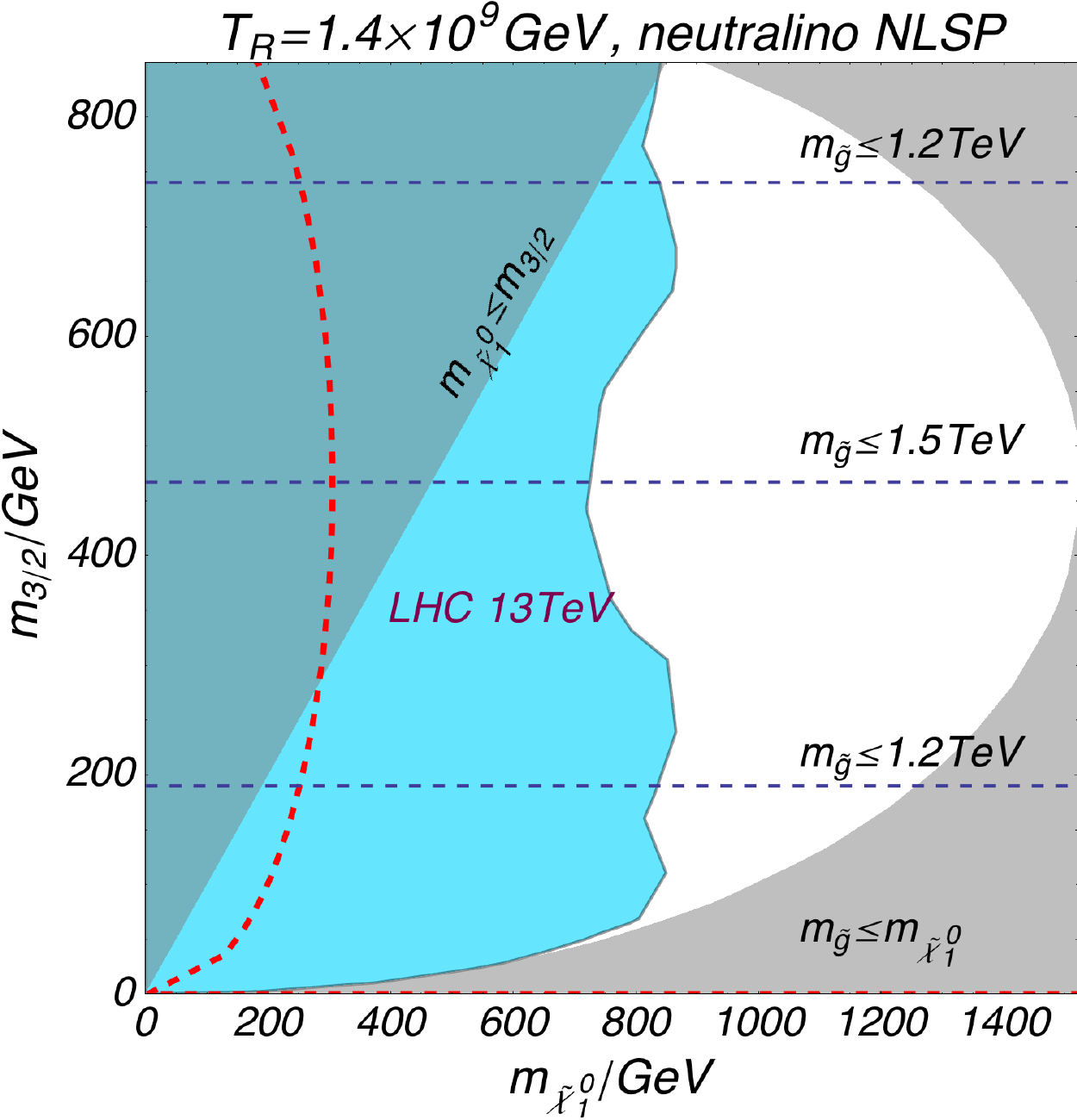}
 \end{minipage}
 \hspace{1cm}
 \begin{minipage}{.45\linewidth}
  \includegraphics[width=\linewidth]{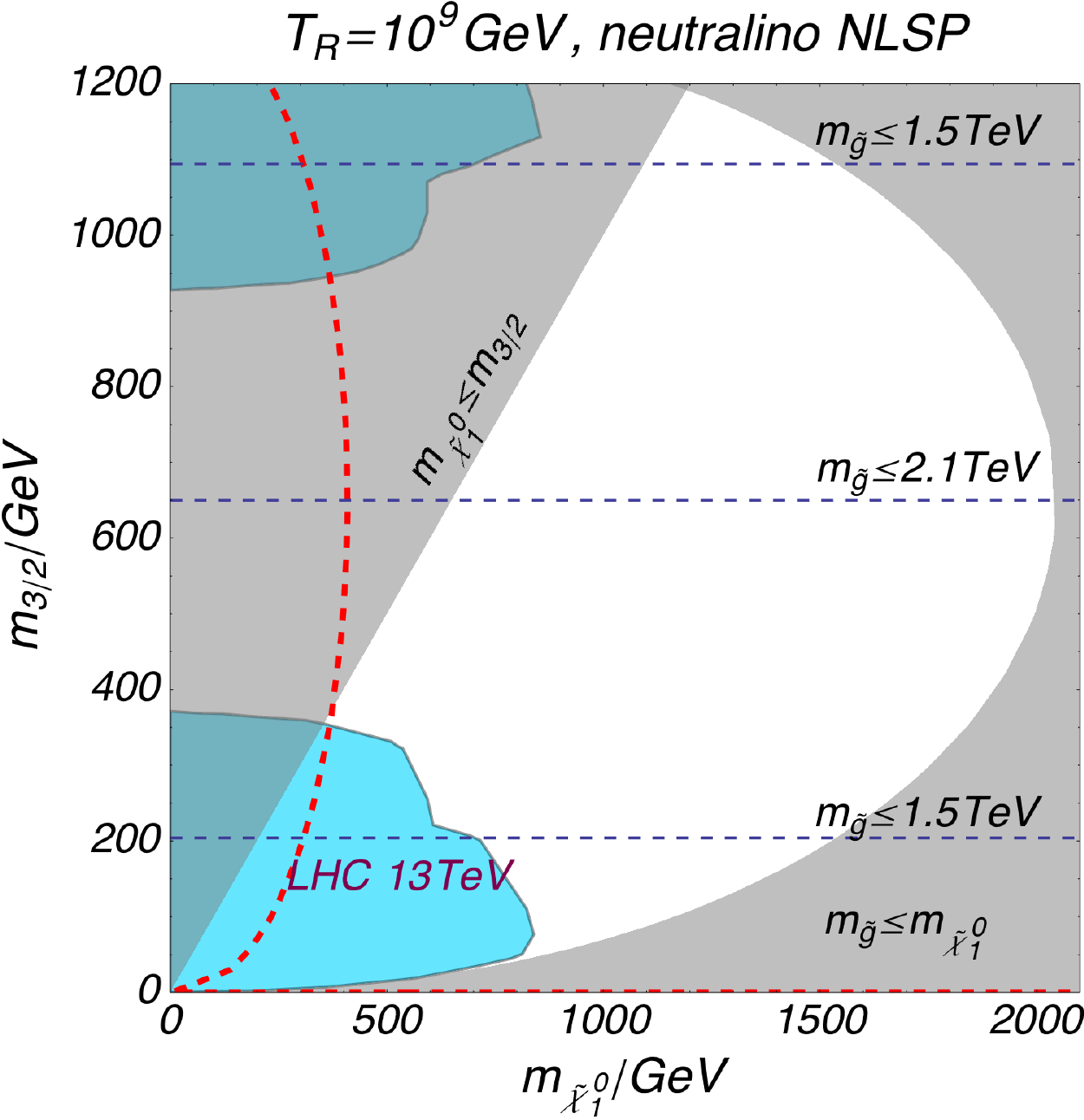}
 \end{minipage}
  \end{center}
\caption{\sl \small  
Combined constraints  for the neutralino NLSP.
The reheating temperature is assumed to be $T_R = 1.4\times10^9$\,GeV (left) and $T_R = 10^9$\,GeV (right).
The gray shaded regions are excluded where the gravitino is no more the LSP.
The blue shaded regions are excluded by the missing momentum searches \cite{ATLAS:2016kts,CMS:2016mwj}.
The GUT relation of the gaugino mass can be satisfied in the left side of the red dashed line.
The horizontal dashed lines show the upper limit on the gluino mass for a given gravitino mass shown in Fig.\,\ref{fig:m32_mglui}.}
\label{fig:_mneu_m32}
\end{figure}

Now, let us combine the constraints  from the gravitino abundance 
in Fig.\,\ref{fig:m32_mglui} with the constraints in Fig.\,\ref{fig:collider} from the collider searches.%
\footnote{In our analysis, we require that the dark matter density is dominated by the gravitino density.
Hence, we assume that the gluino mass should lie on the upper limit in Fig.\,\ref{fig:m32_mglui} for a given gravitino mass.}
In Fig.\,\ref{fig:_mneu_m32}, we show the constraints in the case of the neutralino NLSP
on the $(m_{\tilde \chi_1^0}, m_{3/2})$ plane.
The figure shows that large portion of the parameter region has been excluded by the LHC constraints
for successful leptogenesis, i.e. $T_R \gtrsim 1.4\times 10^9$\,GeV.
Even for somewhat relaxed requirement, $T_R\gtrsim 10^9$\,GeV, some portion of the parameter region has been excluded
by the LHC results.
The remaining allowed region will be tested for $300$\,fb$^{-1}$ of  integrated luminosity at $14$\,TeV
which reaches to $m_{\tilde g}\simeq 2.8$\,TeV~\cite{Cohen:2013zla}.
If we assume the GUT relation to the gaugino masses, the parameter region has been excluded 
even for somewhat lower reheating temperature  $T_R\gtrsim 10^9$\,GeV.

\begin{figure}[t]
\begin{center}
\begin{minipage}{.45\linewidth}
  \includegraphics[width=\linewidth]{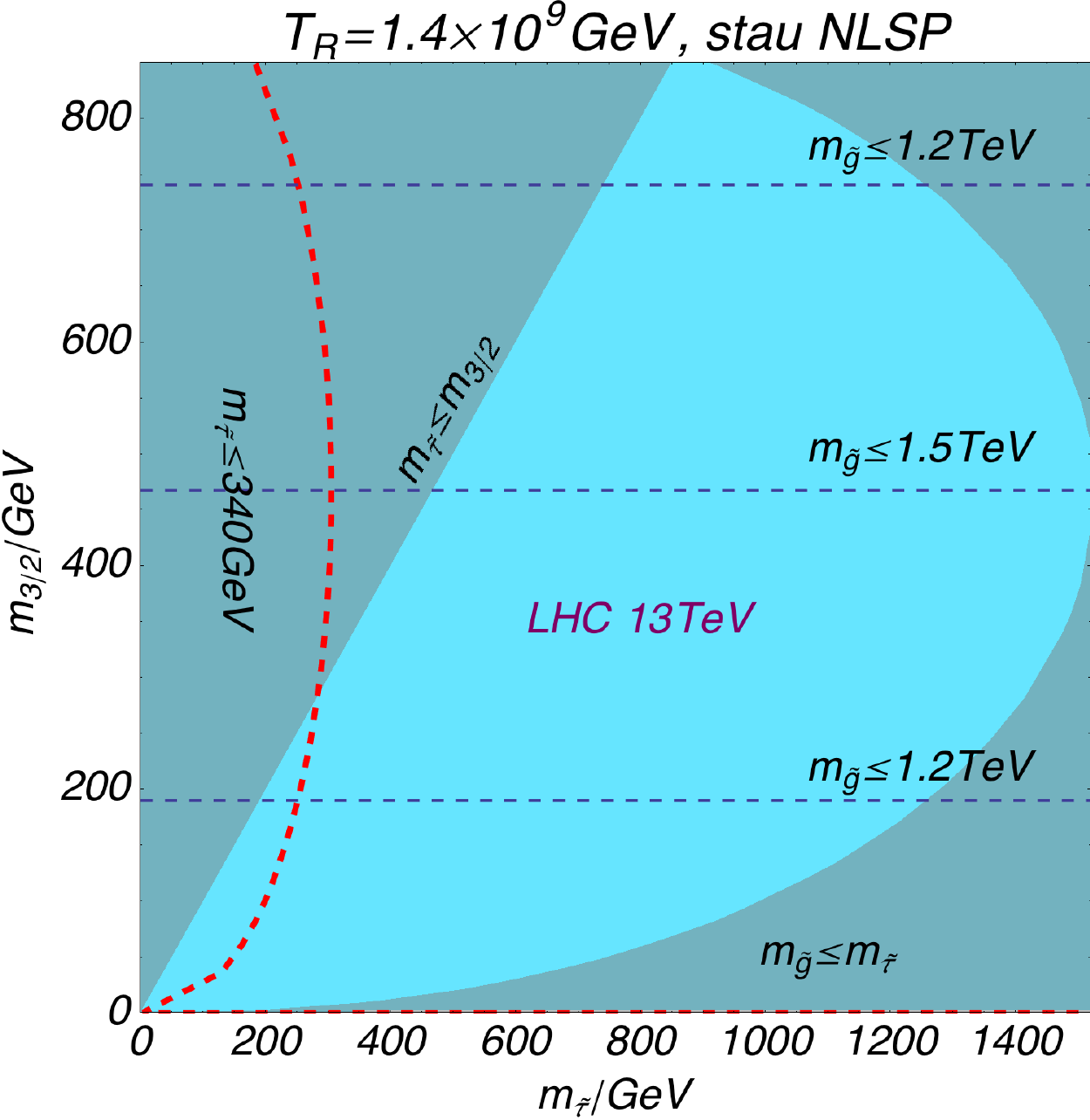}
 \end{minipage}
 \hspace{1cm}
 \begin{minipage}{.45\linewidth}
  \includegraphics[width=\linewidth]{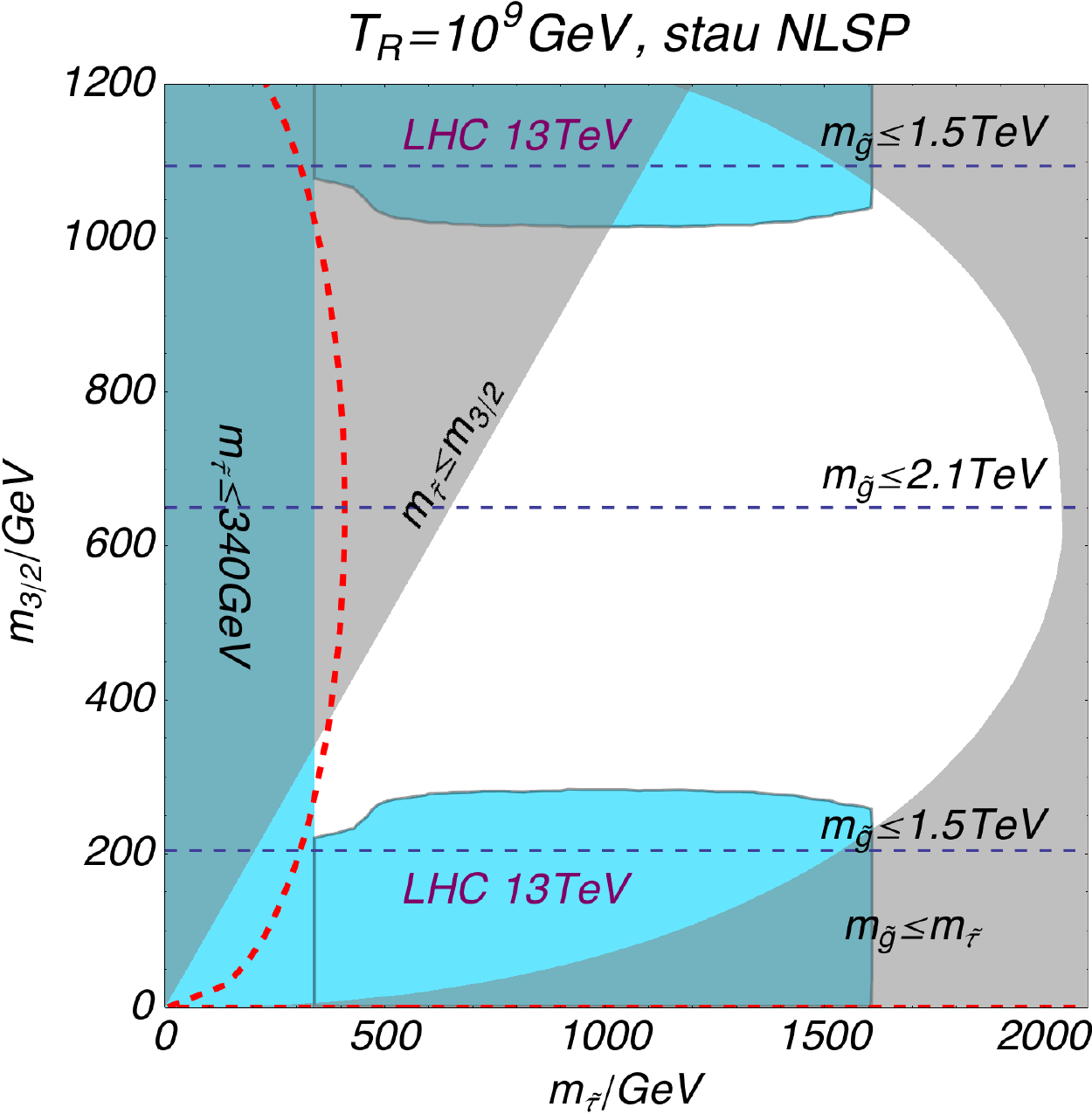}
 \end{minipage}
  \end{center}
\caption{\sl \small  
The same  with \ref{fig:_mneu_m32} but for the stau NLSP.
}
\label{fig:_stau_m32}
\end{figure}

In Fig.\,\ref{fig:_stau_m32}, we also show the combined constraints for the stau NLSP.
The figure shows that all the parameter region has been excluded by the LHC constraints
for $T_R \gtrsim 1.4\times 10^9$\,GeV.
For a relaxed requirement, $T_R \gtrsim 10^9$\,GeV, on the other hand, there 
remains some allowed region, which can be also tested by further data taking.

So far, we have discussed the cases with the neutralino NLSP and the stau NLSP.
Before closing this section, let us comment on  other candidates for the NLSP.
When the gluino is the NLSP, it is again stable inside the detectors, and is hadronized
with the SM quarks to form the so-called $R$-hadrons~\cite{Farrar:1997ns,Kraan:2005ji}.
The $R$-hadrons are charged unless the gluinos are bounded with the gluons,
and the  charged $R$-hadrons can be searched for as long-lived charged particles.
So far, the CMS collaboration has excluded the gluino mass below $1.5$\,TeV at 95\% CL 
when the 50\% of the $R$-hadrons are assumed to be charged~\cite{CMS:2015kdx}.%
\footnote{When the 90\% of the $R$-hadrons are assumed to be charged, the constraints becomes $m_{\tilde g} < 1.59$\,TeV.}
Thus, by comparing with the upper limit on the gluino mass in Fig.\,\ref{fig:m32_mglui},
we find that the gravitino dark matter cannot be consistent with the thermal leptogenesis as in the case with the stau NLSP,
unless the required reheating temperature is somewhat lowered.

The CMS collaboration also puts constraints on the production cross section of the charged $R$-hadron assuming a stable stop particle~\cite{CMS:2015kdx},
which can be applied to the cases of the stop NLSP and  other squark NLSPs.
Since the upper limits on the cross section for a given NLSP mass are tighter than the case of the gluino NLSP,
the lower limits on the gluino mass are tighter for the squark/stop NLSP.
Therefore, we again find that the gravitino dark matter cannot be consistent with the thermal leptogenesis
for these NLSP candidates.

As for the other charged NLSPs such as  selectron/smuon/charginos, the same constraints with the stau NLSP can be applied.
For the sneutrino NLSP, on the other hand, it leaves missing  momentum inside the detector  
as in the case of the neutralino NLSP.
However, we need to perform more detailed analyses including model building to derive constraints,
since the event topologies depend on the decay patterns of the gluinos into the sneutrinos. 
In addition, some careful parameter tunings are required to achieve the sneutrino NLSP in the MSSM. 
From these points of view, we do not pursue this possibility in this paper.

Finally, let us comment on models with a lighter gravitino.
In our discussion, we have assumed that the gravitino is in the hundreds GeV to a few TeV range,
where the NLSP decays outside of the detectors due to a limited size of  $R$-parity violation as in Eq\,(\ref{eq:EGRB}).
If the gravitino mass is a few tens of GeV, on the other hand, the bilinear $R$-parity violation terms can be as large as $\mu'/\mu \sim 10^{-(7-8)}$.%
\footnote{For such a light gravitino, it mainly decays into a pair of a photon and a neutrino. 
The lifetime of such a gravitino is  constrained to be $\tau_{3/2} \gtrsim 10^{29}$\,sec
by the searches for monochromatic gamma-ray line from the Galactic center region~\cite{Ackermann:2015lka}.}
In such cases, the NLSP lifetime can be as short as ${\cal O}(10^{-9})$\,sec in the bi-linear $R$-parity violation~\cite{Buchmuller:2007ui}.
Thus, the neutral NLSP leaves a displaced vertex inside the detectors, and the charged NLSP leaves a kink inside the detectors.%
\footnote{See e.g. \cite{Asai:2011wy,Graham:2012th,Csaki:2015uza} for discussions on the short lived NLSP's.}
For such a light gravitino, however, the gravitino abundance requires $m_{\tilde g}\lesssim 500$\,GeV which are severely
constrained by the searches for multi-track displaced vertices for the neutral NLSP~\cite{TheATLAScollaboration:2013yia,Aad:2015rba}
and by careful reinterpretation~\cite{Evans:2016zau} of the disappearing track searches for the charged NLSP~\cite{Aad:2013yna,CMS:2014gxa}.
We leave detailed analysis for such a light gravitino for future work.

\section{Conclusions and Discussions}
\label{sec:conclusions}
In this paper, we revisited the gravitino dark matter scenario in the presence of the bilinear $R$-parity violating interactions.
In particular, we discussed the consistency with the thermal leptogenesis.
For a high reheating temperature required for the thermal leptogenesis, 
the gravitino dark matter tends to be overproduced, which puts a severe upper limit on the gluino mass.
As a result, we found that a large portion of parameter space has been excluded by the null results of the searches 
for multi-jets with missing momentum at the LHC experiments when the NLSP is assumed to be the neutralino.
For the stau (and other charged slepton) NLSP, on the other hand, more stringent constraints are put 
by the searches for the long-lived charged particles at the LHC experiments. 
As a result, almost all the parameter space has been excluded unless the required reheating temperature 
is somewhat lowered by assuming, for example, a  degenerated right-handed neutrino spectrum.
For the colored NLSP candidates, constraints are tighter than the ones for the stau NLSP, and hence, 
the gravitino dark matter cannot be consistent with thermal leptogenesis in those cases, neither.

It should be noted that the constraints from cosmology are more stringent in the absence of the $R$-parity violation 
since the late-time decay of the NLSP  contributes to the gravitino dark matter abundance~\cite{Fujii:2003nr,Heisig:2013sva,Arvey:2015nra}. 
In addition, the properties of the NLSP are also constrained very severely by the BBN due to 
a long lifetime of the NLSP in the absence of  $R$-parity violation.
As a result, the successful BBN precludes the NLSP candidates other than the charged sleptons or the sneutrinos~\cite{Fujii:2003nr}.
As for the charged sleptons, however, the parameter region has been excluded by the LHC results as discussed in this paper.
The study of the sneutrino NLSP is a future work as mentioned above.

In our discussion, we focused on the bilinear $R$-parity violating interactions which are expected to be dominant 
in wide range of models of spontaneous $R$-parity breaking with the right-handed neutrinos.%
\footnote{See e.g. discussion in \cite{Shirai:2009fq}.}
In fact, once $R$-parity is broken, the linear terms of the right-handed neutrinos, $\varepsilon_{\cancel R} N_R$,
are generically allowed in the superpotential, which leads to $\vev{N}_R \sim \varepsilon_{\cancel R}/M_R$.
Here, $\varepsilon_{\cancel R}$ denotes an $R$-parity violating parameter and $M_R$ the right-handed neutrino mass.
Therefore, the resultant bilinear $R$-parity violating terms are enhanced by $M_R^{-1}$ compared with trilinear 
$R$-parity violating terms which are suppressed not by $M_R$ but by higher cutoff scales such as the Planck scale
or the GUT scale depending on the models.

Nontheless, when $R$-parity violation is dominated by trilinear terms,%
\footnote{See e.g. a model in \cite{Bhattacherjee:2013gr}.}
 the gravitino decays into three SM fermions at a tree-level
and into a pair of a SM boson and a fermion at the one loop level~\cite{Lola:2008bk,Bomark:2009zm}.
In those cases, the upper limits on the sizes of $R$-parity violation from EGRB can be weaker than that in Eq.\,(\ref{eq:EGRB}).
Determination of the upper limits on $R$-parity violation requires more careful analyses in those cases.
If the constraints can be relaxed to the ones in Eq.\,(\ref{eq:cosmology}),%
\footnote{Although this seems difficult unless the gravitino mass is below a few hundred GeV, 
which leads to a severer upper limits on the gluino mass as in Fig.\,\ref{fig:m32_mglui}.}
for example, the NLSP can decay promptly inside the detectors, which relax the LHC constraints.
We leave such studies for future work.

\section*{Acknowledgements}
T.T.Y. thanks Prof. Johannes Blumlein for  hospitality during his stay at DESY in Zeuthen.
This work is supported in part by Grants-in-Aid for Scientific Research from the Ministry of Education, Culture, Sports, Science, and Technology (MEXT) KAKENHI, 
Japan, No. 25105011 and No. 15H05889 (M. I.) as well as No. 26104009 (T. T. Y.); Grant-in-Aid No. 26287039 (M. I. and T. T. Y.) and  No. 16H02176 (T. T. Y.) 
 from the Japan Society for the Promotion of Science (JSPS) KAKENHI; and by the World Premier International Research Center Initiative (WPI), MEXT, Japan (M. I., and T. T. Y.).

\appendix
\section{A model of $R$-parity violation}
\label{sec:model}
\subsection{$R$-parity violation tied with $U(1)_{B-L}$ breaking}
In this appendix, we construct a model where the bilinear $R$-parity violating operators,
\begin{eqnarray}
W = {\mu'}_i H_u L_i  \ ,
\end{eqnarray}
appear in the range of Eq.\,(\ref{eq:Rrange}) which are appropriate for the gravitino dark matter 
in the hundreds GeV to a TeV range.
In particular, we consider a model where the $R$-parity violation is tied to a gauged $U(1)_{B-L}$ breaking
as motivated in $SO(10)$ GUT models~\cite{Buchmuller:2007ui,Schmidt:2010yu}.

Here, we use the $SO(10)$ GUT notation, although it is straight forward to decompose the following discussion
in terms of the MSSM fields.
In the $SO(10)$ GUT models, the quarks and leptons are grouped into ${\bf 16}$ representation, ${\bf 16}_M$,
in conjunction with the right-handed neutrinos.
The Higgs doublets are, on the other hand, grouped into ${\bf 10}$ representation, ${\bf 10}_H$.
In our discussion, we do not specify the mechanisms which explain the doublet-triplet splitting of the Higgs multiplets,
and we assume that only the Higgs doublets in ${\bf 10}_H$ remain below the GUT scale.

In this notation, the MSSM Yukawa couplings are given by,
\begin{eqnarray}
W_{\rm MSSM} = {\bf 10}_H  {\bf 16}_M  {\bf 16}_M \ ,
\end{eqnarray}
where we have suppressed the coefficient and the family indices for simplicity.
To give a large Majorana masses to  the right-handed neutrinos, we need to introduce bilinear terms of
${\bf 16}_M$, which require at least a VEV of ${\bf 126}$ representation. 
However, an introduction of a field in the ${\bf 126}$ representation leads to a blow up of the $SO(10)$ gauge coupling constant just above the GUT scale.
To avoid this problem, we instead break the $U(1)_{B-L}\subset SO(10)$ by a VEV of $\overline{\bf 16}_H$,
\begin{eqnarray}
\vev{\overline{\bf 16}_H} = v_{B-L} \ .
\end{eqnarray}
Here, as in the case of the Higgs doublets, we again assume that only the MSSM singlet in $\overline{\bf 16}_H$. 
With the VEV of $\overline{\bf 16}_H$, the Majorana mass terms are generated from,
\begin{eqnarray}
W_{\rm N_R} = \frac{1}{2M_{\rm PL}}\overline{\bf 16}_H\overline{\bf 16}_H {\bf 16}_M  {\bf 16}_M\ .
\end{eqnarray}
It should be noted no ${\mathbb Z}_2$ subgroup remains unbroken after $U(1)_{B-L}$ breaking.

For a later purpose, we assume that $v_{B-L} = {\cal O}(10^{-(3-4)}) \times M_{\rm PL}$, so that 
the right handed neutrino masses are in the range of
\begin{eqnarray}
M_R \sim \frac{1}{M_{\rm PL}} v_{B-L}^2 \sim10^{-(6-8)}\times M_{\rm PL}\ .
\end{eqnarray}
Then, we aim to construct a model where  the right-handed sneutrinos obtain VEVs
\begin{eqnarray}
\text{$\vev{\tilde{N_R}}$} \simeq \frac{v_{B-L}^3}{M_{\rm PL}^3} \times m_{3/2}\ ,
\end{eqnarray}
while the $R$-parity conserving $\mu$-term is given by $\mu \sim m_{3/2}$.
Once these are achieved, we obtain an appropriate bilinear $R$-parity violating terms,
\begin{eqnarray}
\frac{\mu'}{\mu} \sim \frac{v_{B-L}^3}{M_{\rm PL}^3}\ .
\end{eqnarray}

\begin{table}[t]
\label{tab:R}
\begin{center}
\begin{tabular}{|c|ccccccc|}
\hline\hline
& ${\bf 16}_M$ & ${\bf 10}_H$ &  ${\bf 16}_H$ & $\overline{\bf 16}_H$ & $v_{B-L}$ & $X$ & $m_{3/2}$\\
\hline
$R$ & $1$ & $0$ &  $-1/2$ & $0$ & $-1/4$ & $5/2$ & 2\\
\hline\hline
\end{tabular}
\caption{\small\sl $R$-charges of matter fields, Higgs fields and SO(10) singlets. }
\end{center}
\end{table}%

After the example of the model in \cite{Buchmuller:2007ui}, let us interconnect the $R$-parity violation to the $U(1)_{B-L}$ breaking scale.
First, in order to give a VEV to $\overline{\bf 16}_H$, we consider a superpotential
\begin{eqnarray}
\label{eq:super}
W = {\bf 10}_H  {\bf 16}_M  {\bf 16}_M   +  \overline{\bf 16}_H \overline{\bf 16}_H  {\bf 16}_M  {\bf 16}_M  + X ( {\bf 16}_H  \overline{\bf 16}_H - v_{B-L}^2)\ ,
\end{eqnarray}
where ${\bf 16}_H$ is a newly introduced ${\bf 16}$ representation and $X$ is an $SO(10)$ singlet.
Hereafter, we take the unit of $M_{\rm PL}= 1$.
The first term of Eq.\,(\ref{eq:super}) again denotes the MSSM Yukawa interaction, and the second term the Majorana mass term of the right-handed neutrinos.
By the assumption of the sprit multiplet, ${\bf 16}_H$ and $\overline{\bf 16}_H$  contain the MSSM singlets only which are absorbed 
into $U(1)_{B-L}$ gauge multiplet once they obtain the vacuum expectation value.

In order to avoid too large $R$-parity violations, we forbid the following operators
\begin{eqnarray}
W &=& {\bf 16}_M  \overline {\bf 16}_H \ , \cr
W &=&{\bf 10}_H {\bf 16}_M  {\bf 16}_H  \ , \cr
W &=& {\bf 16}_H  {\bf 16}_M  {\bf 16}_M  {\bf 16}_M \ .
\end{eqnarray}
For that purpose, we consider a continuos $R$-symmetry broken by a spurion $v_{B-L}$ 
(see more discussions in the next subsection). 
In Tab.\,\ref{tab:R}, we show the $R$-charge assignment which forbids the above operators.

A notable feature of the $R$-charge assignment in Tab.\,\ref{tab:R} is that it allows a K\"ahler potential,
\begin{eqnarray}
K = v_{B-L}^4   {\bf 16}_M  \overline {\bf 16}_H\ .
\end{eqnarray}
which leads to a linear term of the right-handed neutrinos%
\footnote{If we regard $m_{3/2}$, we may directly write down this term.} 
\begin{eqnarray}
W\sim m_{3/2} v_{B-L}^5 N_R \ .
\end{eqnarray}
Thus, by combined with the Majorana mass term,  the right-handed sneutrinos obtain VEVs,
\begin{eqnarray}
\vev{\tilde{N_R}} \simeq {v_{B-L}^3} \times m_{3/2}\ ,
\end{eqnarray}
which generate the bilinear $R$-parity violation through the first term of Eq.\,(\ref{eq:super}),
\begin{eqnarray}
\mu' = v_{B-L}^3 m_{3/2}\ .
\end{eqnarray} 
The $R$-symmetric $\mu$-term is, on the other hand, given by
\begin{eqnarray}
W \sim  m_{3/2}{\bf 10}_H {\bf 10}_H\ .
\end{eqnarray}
Therefore, we find that the bilinear $R$-parity violation are given by,%
\footnote{The actual $R$-parity violating bilinear terms are multiplied by the neutrino Yukawa couplings.}
\begin{eqnarray}
\frac{\mu'}{\mu} \sim v_{B-L}^3 \ .
\end{eqnarray}

\subsection{Model with discrete $R$-symmetry}
In the above example, we made use of a continuous $R$-symmetry which is broken by the spurion $v_{B-L}$.
In this subsection, we discuss a model where $v_{B-L}$ is dynamical.
For that purpose, we consider  $SU(5)$ gauge theory with four-flavor of vector-like pairs of fundamental representation $(Q,\bar{Q})$,
and replace $v_{B-L}^2$ to a composite operator $(Q\bar{Q})$.
Then, the superpotential in Eq.\,(\ref{eq:super}) is rewritten by,
\begin{eqnarray}
W = {\bf 10}_H  {\bf 16}_M  {\bf 16}_M   +   \overline{\bf 16}_H \overline{\bf 16}_H  {\bf 16}_M  {\bf 16}_M  + X ( {\bf 16}_H  \overline{\bf 16}_H - Q\bar Q)\ ,
\end{eqnarray}
where the $R$-charge assignment is given in Tab.\,\ref{tab:R2}.%
\footnote{This charge assignment is free from the $SU(5)$ anomaly.}
Since the $SU(5)$ gauge theory with four-flavor does not have a vacuum, we add explicit mass terms
\begin{eqnarray}
W = {\bf 10}_H  {\bf 16}_M  {\bf 16}_M   +   \overline{\bf 16}_H \overline{\bf 16}_H  {\bf 16}_M  {\bf 16}_M  + X ( {\bf 16}_H  \overline{\bf 16}_H - Q\bar Q) + m_Q Q\bar{Q}\ .
\end{eqnarray}
With the explicit mass term,  the $R$-symmetry is explicitly broken down to a discrete ${\mathbb Z}_{10 R}$ symmetry
whose charge assignment is given in the second line of Tab.\,\ref{tab:R2}.%
\footnote{As we will see, we require $m_Q \ll 1$. For that purpose, we assume that $m_Q$, $X$, ${\bf 16}_H$ and $Q\bar{Q}$
are charged under some additional discrete symmetry.}

\begin{table}[t]
\label{tab:R2}
\begin{center}
\begin{tabular}{|c|ccccccc|}
\hline\hline
& ${\bf 16}_M$ & ${\bf 10}_H$ &  ${\bf 16}_H$ & $\overline{\bf 16}_H$ &  $Q$ & $\bar{Q}$ & $X$ \\
\hline
$R$ & $1$ & $0$ &  $-1/2$ & $0$ & $-1/4$& $-1/4$& $5/2$  \\
\hline
${\mathbb Z}_{10R}$ & $-4$ & $0$ &  $2$ & $0$ & $1$& $1$& $0$  \\
\hline\hline
\end{tabular}
\caption{\small\sl $R$-charges of matter fields, Higgs fields and SO(10) singlets.
We also show the charge assignment of ${\mathbb Z}_{10R}$ with which the terms with charges $-8$ (mod $10$) 
are allowed in the superpotential.}
\end{center}
\end{table}%

Below the dynamical scale of $SU(5)$, non-perturbative potential is generated~\cite{Affleck:1983mk} 
\begin{eqnarray}
W &=& {\bf 10}_H  {\bf 16}_M  {\bf 16}_M   +   \overline{\bf 16}_H \overline{\bf 16}_H  {\bf 16}_M  {\bf 16}_M  + X ( {\bf 16}_H  \overline{\bf 16}_H - Q\bar Q)\nonumber\\
&& + m_Q Q\bar{Q}
+ \frac{\L^{11}}{\det{Q\bar{Q}}}\ ,
\end{eqnarray}
where $\Lambda$ denotes the dynamical scale of $SU(5)$.
As a result, $Q\bar{Q}$ obtains a VEV, which provides the spurion in the previous section
\begin{eqnarray}
v_{B-L}^2\sim \vev{Q\bar{Q}} \sim \left(\frac{\L^{11}}{m_Q}\right)^{1/5}\ .
\end{eqnarray}
Thus, by arranging 
\begin{eqnarray}
\left(\frac{\L^{11}}{m_Q}\right)^{1/5} = {\cal O}(10^{-6})\ ,
\end{eqnarray}
we can provide an appropriate suprion $ v_{B-L} = {\cal O}(10^{-3})$.
Furthermore, we can also provide an appropriate size of the gravition mass by taking 
\begin{eqnarray}
m_Q \simeq 10^{-9}\ , \quad \L \simeq 10^{-3.5} \ ,
\end{eqnarray}
which leads to an appropriate VEV of the superpotential simultaneously
\begin{eqnarray}
m_{3/2} = \vev{W} \sim m_Q   \vev{Q\bar{Q}} \simeq 10^{-15}\ .
\end{eqnarray}


\bibliography{draft_resubmit.bbl}

\begin{thebibliography}{90}%
\makeatletter
\providecommand \@ifxundefined [1]{%
 \@ifx{#1\undefined}
}%
\providecommand \@ifnum [1]{%
 \ifnum #1\expandafter \@firstoftwo
 \else \expandafter \@secondoftwo
 \fi
}%
\providecommand \@ifx [1]{%
 \ifx #1\expandafter \@firstoftwo
 \else \expandafter \@secondoftwo
 \fi
}%
\providecommand \natexlab [1]{#1}%
\providecommand \enquote  [1]{``#1''}%
\providecommand \bibnamefont  [1]{#1}%
\providecommand \bibfnamefont [1]{#1}%
\providecommand \citenamefont [1]{#1}%
\providecommand \href@noop [0]{\@secondoftwo}%
\providecommand \href [0]{\begingroup \@sanitize@url \@href}%
\providecommand \@href[1]{\@@startlink{#1}\@@href}%
\providecommand \@@href[1]{\endgroup#1\@@endlink}%
\providecommand \@sanitize@url [0]{\catcode `\\12\catcode `\$12\catcode
  `\&12\catcode `\#12\catcode `\^12\catcode `\_12\catcode `\%12\relax}%
\providecommand \@@startlink[1]{}%
\providecommand \@@endlink[0]{}%
\providecommand \url  [0]{\begingroup\@sanitize@url \@url }%
\providecommand \@url [1]{\endgroup\@href {#1}{\urlprefix }}%
\providecommand \urlprefix  [0]{URL }%
\providecommand \Eprint [0]{\href }%
\providecommand \doibase [0]{http://dx.doi.org/}%
\providecommand \selectlanguage [0]{\@gobble}%
\providecommand \bibinfo  [0]{\@secondoftwo}%
\providecommand \bibfield  [0]{\@secondoftwo}%
\providecommand \translation [1]{[#1]}%
\providecommand \BibitemOpen [0]{}%
\providecommand \bibitemStop [0]{}%
\providecommand \bibitemNoStop [0]{.\EOS\space}%
\providecommand \EOS [0]{\spacefactor3000\relax}%
\providecommand \BibitemShut  [1]{\csname bibitem#1\endcsname}%
\let\auto@bib@innerbib\@empty
\bibitem [{\citenamefont {Fayet}(1976)}]{Fayet:1976et}%
  \BibitemOpen
  \bibfield  {author} {\bibinfo {author} {\bibfnamefont {P.}~\bibnamefont
  {Fayet}},\ }\href {\doibase 10.1016/0370-2693(76)90319-1} {\bibfield
  {journal} {\bibinfo  {journal} {Phys. Lett.}\ }\textbf {\bibinfo {volume}
  {B64}},\ \bibinfo {pages} {159} (\bibinfo {year} {1976})}\BibitemShut
  {NoStop}%
\bibitem [{\citenamefont {Fayet}(1977)}]{Fayet:1977yc}%
  \BibitemOpen
  \bibfield  {author} {\bibinfo {author} {\bibfnamefont {P.}~\bibnamefont
  {Fayet}},\ }\href {\doibase 10.1016/0370-2693(77)90852-8} {\bibfield
  {journal} {\bibinfo  {journal} {Phys. Lett.}\ }\textbf {\bibinfo {volume}
  {B69}},\ \bibinfo {pages} {489} (\bibinfo {year} {1977})}\BibitemShut
  {NoStop}%
\bibitem [{\citenamefont {Hall}\ and\ \citenamefont
  {Suzuki}(1984)}]{Hall:1983id}%
  \BibitemOpen
  \bibfield  {author} {\bibinfo {author} {\bibfnamefont {L.~J.}\ \bibnamefont
  {Hall}}\ and\ \bibinfo {author} {\bibfnamefont {M.}~\bibnamefont {Suzuki}},\
  }\href {\doibase 10.1016/0550-3213(84)90513-3} {\bibfield  {journal}
  {\bibinfo  {journal} {Nucl. Phys.}\ }\textbf {\bibinfo {volume} {B231}},\
  \bibinfo {pages} {419} (\bibinfo {year} {1984})}\BibitemShut {NoStop}%
\bibitem [{\citenamefont {Hawking}(1987)}]{Hawking:1987mz}%
  \BibitemOpen
  \bibfield  {author} {\bibinfo {author} {\bibfnamefont {S.~W.}\ \bibnamefont
  {Hawking}},\ }\bibfield  {booktitle} {\emph {\bibinfo {booktitle} {{Moscow
  Quantum Grav.1987:0125}}},\ }\href {\doibase 10.1016/0370-2693(87)90028-1}
  {\bibfield  {journal} {\bibinfo  {journal} {Phys. Lett.}\ }\textbf {\bibinfo
  {volume} {B195}},\ \bibinfo {pages} {337} (\bibinfo {year}
  {1987})}\BibitemShut {NoStop}%
\bibitem [{\citenamefont {Lavrelashvili}\ \emph {et~al.}(1987)\citenamefont
  {Lavrelashvili}, \citenamefont {Rubakov},\ and\ \citenamefont
  {Tinyakov}}]{Lavrelashvili:1987jg}%
  \BibitemOpen
  \bibfield  {author} {\bibinfo {author} {\bibfnamefont {G.~V.}\ \bibnamefont
  {Lavrelashvili}}, \bibinfo {author} {\bibfnamefont {V.~A.}\ \bibnamefont
  {Rubakov}}, \ and\ \bibinfo {author} {\bibfnamefont {P.~G.}\ \bibnamefont
  {Tinyakov}},\ }\href@noop {} {\bibfield  {journal} {\bibinfo  {journal} {JETP
  Lett.}\ }\textbf {\bibinfo {volume} {46}},\ \bibinfo {pages} {167} (\bibinfo
  {year} {1987})},\ \bibinfo {note} {[Pisma Zh. Eksp. Teor.
  Fiz.46,134(1987)]}\BibitemShut {NoStop}%
\bibitem [{\citenamefont {Giddings}\ and\ \citenamefont
  {Strominger}(1988)}]{Giddings:1988cx}%
  \BibitemOpen
  \bibfield  {author} {\bibinfo {author} {\bibfnamefont {S.~B.}\ \bibnamefont
  {Giddings}}\ and\ \bibinfo {author} {\bibfnamefont {A.}~\bibnamefont
  {Strominger}},\ }\href {\doibase 10.1016/0550-3213(88)90109-5} {\bibfield
  {journal} {\bibinfo  {journal} {Nucl. Phys.}\ }\textbf {\bibinfo {volume}
  {B307}},\ \bibinfo {pages} {854} (\bibinfo {year} {1988})}\BibitemShut
  {NoStop}%
\bibitem [{\citenamefont {Coleman}(1988)}]{Coleman:1988tj}%
  \BibitemOpen
  \bibfield  {author} {\bibinfo {author} {\bibfnamefont {S.~R.}\ \bibnamefont
  {Coleman}},\ }\href {\doibase 10.1016/0550-3213(88)90097-1} {\bibfield
  {journal} {\bibinfo  {journal} {Nucl. Phys.}\ }\textbf {\bibinfo {volume}
  {B310}},\ \bibinfo {pages} {643} (\bibinfo {year} {1988})}\BibitemShut
  {NoStop}%
\bibitem [{\citenamefont {Gilbert}(1989)}]{Gilbert:1989nq}%
  \BibitemOpen
  \bibfield  {author} {\bibinfo {author} {\bibfnamefont {G.}~\bibnamefont
  {Gilbert}},\ }\href {\doibase 10.1016/0550-3213(89)90097-7} {\bibfield
  {journal} {\bibinfo  {journal} {Nucl. Phys.}\ }\textbf {\bibinfo {volume}
  {B328}},\ \bibinfo {pages} {159} (\bibinfo {year} {1989})}\BibitemShut
  {NoStop}%
\bibitem [{\citenamefont {Banks}\ and\ \citenamefont
  {Seiberg}(2011)}]{Banks:2010zn}%
  \BibitemOpen
  \bibfield  {author} {\bibinfo {author} {\bibfnamefont {T.}~\bibnamefont
  {Banks}}\ and\ \bibinfo {author} {\bibfnamefont {N.}~\bibnamefont
  {Seiberg}},\ }\href {\doibase 10.1103/PhysRevD.83.084019} {\bibfield
  {journal} {\bibinfo  {journal} {Phys. Rev.}\ }\textbf {\bibinfo {volume}
  {D83}},\ \bibinfo {pages} {084019} (\bibinfo {year} {2011})},\ \Eprint
  {http://arxiv.org/abs/1011.5120} {arXiv:1011.5120 [hep-th]} \BibitemShut
  {NoStop}%
\bibitem [{\citenamefont {Krauss}\ and\ \citenamefont
  {Wilczek}(1989)}]{Krauss:1988zc}%
  \BibitemOpen
  \bibfield  {author} {\bibinfo {author} {\bibfnamefont {L.~M.}\ \bibnamefont
  {Krauss}}\ and\ \bibinfo {author} {\bibfnamefont {F.}~\bibnamefont
  {Wilczek}},\ }\href {\doibase 10.1103/PhysRevLett.62.1221} {\bibfield
  {journal} {\bibinfo  {journal} {Phys. Rev. Lett.}\ }\textbf {\bibinfo
  {volume} {62}},\ \bibinfo {pages} {1221} (\bibinfo {year}
  {1989})}\BibitemShut {NoStop}%
\bibitem [{\citenamefont {Preskill}\ and\ \citenamefont
  {Krauss}(1990)}]{Preskill:1990bm}%
  \BibitemOpen
  \bibfield  {author} {\bibinfo {author} {\bibfnamefont {J.}~\bibnamefont
  {Preskill}}\ and\ \bibinfo {author} {\bibfnamefont {L.~M.}\ \bibnamefont
  {Krauss}},\ }\href {\doibase 10.1016/0550-3213(90)90262-C} {\bibfield
  {journal} {\bibinfo  {journal} {Nucl. Phys.}\ }\textbf {\bibinfo {volume}
  {B341}},\ \bibinfo {pages} {50} (\bibinfo {year} {1990})}\BibitemShut
  {NoStop}%
\bibitem [{\citenamefont {Preskill}\ \emph {et~al.}(1991)\citenamefont
  {Preskill}, \citenamefont {Trivedi}, \citenamefont {Wilczek},\ and\
  \citenamefont {Wise}}]{Preskill:1991kd}%
  \BibitemOpen
  \bibfield  {author} {\bibinfo {author} {\bibfnamefont {J.}~\bibnamefont
  {Preskill}}, \bibinfo {author} {\bibfnamefont {S.~P.}\ \bibnamefont
  {Trivedi}}, \bibinfo {author} {\bibfnamefont {F.}~\bibnamefont {Wilczek}}, \
  and\ \bibinfo {author} {\bibfnamefont {M.~B.}\ \bibnamefont {Wise}},\ }\href
  {\doibase 10.1016/0550-3213(91)90241-O} {\bibfield  {journal} {\bibinfo
  {journal} {Nucl. Phys.}\ }\textbf {\bibinfo {volume} {B363}},\ \bibinfo
  {pages} {207} (\bibinfo {year} {1991})}\BibitemShut {NoStop}%
\bibitem [{\citenamefont {Banks}\ and\ \citenamefont
  {Dine}(1992)}]{Banks:1991xj}%
  \BibitemOpen
  \bibfield  {author} {\bibinfo {author} {\bibfnamefont {T.}~\bibnamefont
  {Banks}}\ and\ \bibinfo {author} {\bibfnamefont {M.}~\bibnamefont {Dine}},\
  }\href {\doibase 10.1103/PhysRevD.45.1424} {\bibfield  {journal} {\bibinfo
  {journal} {Phys. Rev.}\ }\textbf {\bibinfo {volume} {D45}},\ \bibinfo {pages}
  {1424} (\bibinfo {year} {1992})},\ \Eprint
  {http://arxiv.org/abs/hep-th/9109045} {arXiv:hep-th/9109045 [hep-th]}
  \BibitemShut {NoStop}%
\bibitem [{\citenamefont {Dimopoulos}\ and\ \citenamefont
  {Georgi}(1981)}]{Dimopoulos:1981zb}%
  \BibitemOpen
  \bibfield  {author} {\bibinfo {author} {\bibfnamefont {S.}~\bibnamefont
  {Dimopoulos}}\ and\ \bibinfo {author} {\bibfnamefont {H.}~\bibnamefont
  {Georgi}},\ }\href {\doibase 10.1016/0550-3213(81)90522-8} {\bibfield
  {journal} {\bibinfo  {journal} {Nucl. Phys.}\ }\textbf {\bibinfo {volume}
  {B193}},\ \bibinfo {pages} {150} (\bibinfo {year} {1981})}\BibitemShut
  {NoStop}%
\bibitem [{\citenamefont {Weinberg}(1982)}]{Weinberg:1981wj}%
  \BibitemOpen
  \bibfield  {author} {\bibinfo {author} {\bibfnamefont {S.}~\bibnamefont
  {Weinberg}},\ }\href {\doibase 10.1103/PhysRevD.26.287} {\bibfield  {journal}
  {\bibinfo  {journal} {Phys. Rev.}\ }\textbf {\bibinfo {volume} {D26}},\
  \bibinfo {pages} {287} (\bibinfo {year} {1982})}\BibitemShut {NoStop}%
\bibitem [{\citenamefont {Sakai}\ and\ \citenamefont
  {Yanagida}(1982)}]{Sakai:1981pk}%
  \BibitemOpen
  \bibfield  {author} {\bibinfo {author} {\bibfnamefont {N.}~\bibnamefont
  {Sakai}}\ and\ \bibinfo {author} {\bibfnamefont {T.}~\bibnamefont
  {Yanagida}},\ }\href {\doibase 10.1016/0550-3213(82)90457-6} {\bibfield
  {journal} {\bibinfo  {journal} {Nucl. Phys.}\ }\textbf {\bibinfo {volume}
  {B197}},\ \bibinfo {pages} {533} (\bibinfo {year} {1982})}\BibitemShut
  {NoStop}%
\bibitem [{\citenamefont {Dimopoulos}\ \emph {et~al.}(1982)\citenamefont
  {Dimopoulos}, \citenamefont {Raby},\ and\ \citenamefont
  {Wilczek}}]{Dimopoulos:1981dw}%
  \BibitemOpen
  \bibfield  {author} {\bibinfo {author} {\bibfnamefont {S.}~\bibnamefont
  {Dimopoulos}}, \bibinfo {author} {\bibfnamefont {S.}~\bibnamefont {Raby}}, \
  and\ \bibinfo {author} {\bibfnamefont {F.}~\bibnamefont {Wilczek}},\ }\href
  {\doibase 10.1016/0370-2693(82)90313-6} {\bibfield  {journal} {\bibinfo
  {journal} {Phys. Lett.}\ }\textbf {\bibinfo {volume} {B112}},\ \bibinfo
  {pages} {133} (\bibinfo {year} {1982})}\BibitemShut {NoStop}%
\bibitem [{\citenamefont {Yanagida}(1979)}]{Yanagida:1979as}%
  \BibitemOpen
  \bibfield  {author} {\bibinfo {author} {\bibfnamefont {T.}~\bibnamefont
  {Yanagida}},\ }\bibfield  {booktitle} {\emph {\bibinfo {booktitle}
  {{Proceedings: Workshop on the Unified Theories and the Baryon Number in the
  Universe: Tsukuba, Japan, February 13-14, 1979}}},\ }\href@noop {} {\bibfield
   {journal} {\bibinfo  {journal} {Conf. Proc.}\ }\textbf {\bibinfo {volume}
  {C7902131}},\ \bibinfo {pages} {95} (\bibinfo {year} {1979})}\BibitemShut
  {NoStop}%
\bibitem [{\citenamefont {Ramond}(1979)}]{Ramond:1979py}%
  \BibitemOpen
  \bibfield  {author} {\bibinfo {author} {\bibfnamefont {P.}~\bibnamefont
  {Ramond}},\ }in\ \href@noop {} {\emph {\bibinfo {booktitle} {{International
  Symposium on Fundamentals of Quantum Theory and Quantum Field Theory Palm
  Coast, Florida, February 25-March 2, 1979}}}}\ (\bibinfo {year} {1979})\ pp.\
  \bibinfo {pages} {265--280},\ \Eprint {http://arxiv.org/abs/hep-ph/9809459}
  {arXiv:hep-ph/9809459 [hep-ph]} \BibitemShut {NoStop}%
\bibitem [{\citenamefont {Minkowski}(1977)}]{Minkowski:1977sc}%
  \BibitemOpen
  \bibfield  {author} {\bibinfo {author} {\bibfnamefont {P.}~\bibnamefont
  {Minkowski}},\ }\href {\doibase 10.1016/0370-2693(77)90435-X} {\bibfield
  {journal} {\bibinfo  {journal} {Phys. Lett.}\ }\textbf {\bibinfo {volume}
  {B67}},\ \bibinfo {pages} {421} (\bibinfo {year} {1977})}\BibitemShut
  {NoStop}%
\bibitem [{\citenamefont {Watari}(2015)}]{Watari:2015ysa}%
  \BibitemOpen
  \bibfield  {author} {\bibinfo {author} {\bibfnamefont {T.}~\bibnamefont
  {Watari}},\ }\href {\doibase 10.1007/JHEP11(2015)065} {\bibfield  {journal}
  {\bibinfo  {journal} {JHEP}\ }\textbf {\bibinfo {volume} {11}},\ \bibinfo
  {pages} {065} (\bibinfo {year} {2015})},\ \Eprint
  {http://arxiv.org/abs/1506.08433} {arXiv:1506.08433 [hep-th]} \BibitemShut
  {NoStop}%
\bibitem [{\citenamefont {Buchmuller}\ \emph {et~al.}(2007)\citenamefont
  {Buchmuller}, \citenamefont {Covi}, \citenamefont {Hamaguchi}, \citenamefont
  {Ibarra},\ and\ \citenamefont {Yanagida}}]{Buchmuller:2007ui}%
  \BibitemOpen
  \bibfield  {author} {\bibinfo {author} {\bibfnamefont {W.}~\bibnamefont
  {Buchmuller}}, \bibinfo {author} {\bibfnamefont {L.}~\bibnamefont {Covi}},
  \bibinfo {author} {\bibfnamefont {K.}~\bibnamefont {Hamaguchi}}, \bibinfo
  {author} {\bibfnamefont {A.}~\bibnamefont {Ibarra}}, \ and\ \bibinfo {author}
  {\bibfnamefont {T.}~\bibnamefont {Yanagida}},\ }\href {\doibase
  10.1088/1126-6708/2007/03/037} {\bibfield  {journal} {\bibinfo  {journal}
  {JHEP}\ }\textbf {\bibinfo {volume} {03}},\ \bibinfo {pages} {037} (\bibinfo
  {year} {2007})},\ \Eprint {http://arxiv.org/abs/hep-ph/0702184}
  {arXiv:hep-ph/0702184 [HEP-PH]} \BibitemShut {NoStop}%
\bibitem [{\citenamefont {Schmidt}\ \emph {et~al.}(2010)\citenamefont
  {Schmidt}, \citenamefont {Weniger},\ and\ \citenamefont
  {Yanagida}}]{Schmidt:2010yu}%
  \BibitemOpen
  \bibfield  {author} {\bibinfo {author} {\bibfnamefont {J.}~\bibnamefont
  {Schmidt}}, \bibinfo {author} {\bibfnamefont {C.}~\bibnamefont {Weniger}}, \
  and\ \bibinfo {author} {\bibfnamefont {T.~T.}\ \bibnamefont {Yanagida}},\
  }\href {\doibase 10.1103/PhysRevD.82.103517} {\bibfield  {journal} {\bibinfo
  {journal} {Phys. Rev.}\ }\textbf {\bibinfo {volume} {D82}},\ \bibinfo {pages}
  {103517} (\bibinfo {year} {2010})},\ \Eprint {http://arxiv.org/abs/1008.0398}
  {arXiv:1008.0398 [hep-ph]} \BibitemShut {NoStop}%
\bibitem [{\citenamefont {Takayama}\ and\ \citenamefont
  {Yamaguchi}(2000)}]{Takayama:2000uz}%
  \BibitemOpen
  \bibfield  {author} {\bibinfo {author} {\bibfnamefont {F.}~\bibnamefont
  {Takayama}}\ and\ \bibinfo {author} {\bibfnamefont {M.}~\bibnamefont
  {Yamaguchi}},\ }\href {\doibase 10.1016/S0370-2693(00)00726-7} {\bibfield
  {journal} {\bibinfo  {journal} {Phys. Lett.}\ }\textbf {\bibinfo {volume}
  {B485}},\ \bibinfo {pages} {388} (\bibinfo {year} {2000})},\ \Eprint
  {http://arxiv.org/abs/hep-ph/0005214} {arXiv:hep-ph/0005214 [hep-ph]}
  \BibitemShut {NoStop}%
\bibitem [{\citenamefont {Moreau}\ and\ \citenamefont
  {Chemtob}(2002)}]{Moreau:2001sr}%
  \BibitemOpen
  \bibfield  {author} {\bibinfo {author} {\bibfnamefont {G.}~\bibnamefont
  {Moreau}}\ and\ \bibinfo {author} {\bibfnamefont {M.}~\bibnamefont
  {Chemtob}},\ }\href {\doibase 10.1103/PhysRevD.65.024033} {\bibfield
  {journal} {\bibinfo  {journal} {Phys. Rev.}\ }\textbf {\bibinfo {volume}
  {D65}},\ \bibinfo {pages} {024033} (\bibinfo {year} {2002})},\ \Eprint
  {http://arxiv.org/abs/hep-ph/0107286} {arXiv:hep-ph/0107286 [hep-ph]}
  \BibitemShut {NoStop}%
\bibitem [{\citenamefont {Fukugita}\ and\ \citenamefont
  {Yanagida}(1986)}]{Fukugita:1986hr}%
  \BibitemOpen
  \bibfield  {author} {\bibinfo {author} {\bibfnamefont {M.}~\bibnamefont
  {Fukugita}}\ and\ \bibinfo {author} {\bibfnamefont {T.}~\bibnamefont
  {Yanagida}},\ }\href {\doibase 10.1016/0370-2693(86)91126-3} {\bibfield
  {journal} {\bibinfo  {journal} {Phys. Lett.}\ }\textbf {\bibinfo {volume}
  {B174}},\ \bibinfo {pages} {45} (\bibinfo {year} {1986})}\BibitemShut
  {NoStop}%
\bibitem [{\citenamefont {Giudice}\ \emph {et~al.}(2004)\citenamefont
  {Giudice}, \citenamefont {Notari}, \citenamefont {Raidal}, \citenamefont
  {Riotto},\ and\ \citenamefont {Strumia}}]{Giudice:2003jh}%
  \BibitemOpen
  \bibfield  {author} {\bibinfo {author} {\bibfnamefont {G.~F.}\ \bibnamefont
  {Giudice}}, \bibinfo {author} {\bibfnamefont {A.}~\bibnamefont {Notari}},
  \bibinfo {author} {\bibfnamefont {M.}~\bibnamefont {Raidal}}, \bibinfo
  {author} {\bibfnamefont {A.}~\bibnamefont {Riotto}}, \ and\ \bibinfo {author}
  {\bibfnamefont {A.}~\bibnamefont {Strumia}},\ }\href {\doibase
  10.1016/j.nuclphysb.2004.02.019} {\bibfield  {journal} {\bibinfo  {journal}
  {Nucl. Phys.}\ }\textbf {\bibinfo {volume} {B685}},\ \bibinfo {pages} {89}
  (\bibinfo {year} {2004})},\ \Eprint {http://arxiv.org/abs/hep-ph/0310123}
  {arXiv:hep-ph/0310123 [hep-ph]} \BibitemShut {NoStop}%
\bibitem [{\citenamefont {Buchmuller}\ \emph {et~al.}(2005)\citenamefont
  {Buchmuller}, \citenamefont {Peccei},\ and\ \citenamefont
  {Yanagida}}]{Buchmuller:2005eh}%
  \BibitemOpen
  \bibfield  {author} {\bibinfo {author} {\bibfnamefont {W.}~\bibnamefont
  {Buchmuller}}, \bibinfo {author} {\bibfnamefont {R.~D.}\ \bibnamefont
  {Peccei}}, \ and\ \bibinfo {author} {\bibfnamefont {T.}~\bibnamefont
  {Yanagida}},\ }\href {\doibase 10.1146/annurev.nucl.55.090704.151558}
  {\bibfield  {journal} {\bibinfo  {journal} {Ann. Rev. Nucl. Part. Sci.}\
  }\textbf {\bibinfo {volume} {55}},\ \bibinfo {pages} {311} (\bibinfo {year}
  {2005})},\ \Eprint {http://arxiv.org/abs/hep-ph/0502169}
  {arXiv:hep-ph/0502169 [hep-ph]} \BibitemShut {NoStop}%
\bibitem [{\citenamefont {Davidson}\ \emph {et~al.}(2008)\citenamefont
  {Davidson}, \citenamefont {Nardi},\ and\ \citenamefont
  {Nir}}]{Davidson:2008bu}%
  \BibitemOpen
  \bibfield  {author} {\bibinfo {author} {\bibfnamefont {S.}~\bibnamefont
  {Davidson}}, \bibinfo {author} {\bibfnamefont {E.}~\bibnamefont {Nardi}}, \
  and\ \bibinfo {author} {\bibfnamefont {Y.}~\bibnamefont {Nir}},\ }\href
  {\doibase 10.1016/j.physrep.2008.06.002} {\bibfield  {journal} {\bibinfo
  {journal} {Phys. Rept.}\ }\textbf {\bibinfo {volume} {466}},\ \bibinfo
  {pages} {105} (\bibinfo {year} {2008})},\ \Eprint
  {http://arxiv.org/abs/0802.2962} {arXiv:0802.2962 [hep-ph]} \BibitemShut
  {NoStop}%
\bibitem [{\citenamefont {Bolz}\ \emph {et~al.}(2001)\citenamefont {Bolz},
  \citenamefont {Brandenburg},\ and\ \citenamefont {Buchmuller}}]{Bolz:2000fu}%
  \BibitemOpen
  \bibfield  {author} {\bibinfo {author} {\bibfnamefont {M.}~\bibnamefont
  {Bolz}}, \bibinfo {author} {\bibfnamefont {A.}~\bibnamefont {Brandenburg}}, \
  and\ \bibinfo {author} {\bibfnamefont {W.}~\bibnamefont {Buchmuller}},\
  }\href {\doibase 10.1016/S0550-3213(01)00132-8,
  10.1016/j.nuclphysb.2007.09.020} {\bibfield  {journal} {\bibinfo  {journal}
  {Nucl. Phys.}\ }\textbf {\bibinfo {volume} {B606}},\ \bibinfo {pages} {518}
  (\bibinfo {year} {2001})},\ \bibinfo {note} {[Erratum: Nucl.
  Phys.B790,336(2008)]},\ \Eprint {http://arxiv.org/abs/hep-ph/0012052}
  {arXiv:hep-ph/0012052 [hep-ph]} \BibitemShut {NoStop}%
\bibitem [{\citenamefont {Hamaguchi}\ \emph {et~al.}(2009)\citenamefont
  {Hamaguchi}, \citenamefont {Takahashi},\ and\ \citenamefont
  {Yanagida}}]{Hamaguchi:2009sz}%
  \BibitemOpen
  \bibfield  {author} {\bibinfo {author} {\bibfnamefont {K.}~\bibnamefont
  {Hamaguchi}}, \bibinfo {author} {\bibfnamefont {F.}~\bibnamefont
  {Takahashi}}, \ and\ \bibinfo {author} {\bibfnamefont {T.~T.}\ \bibnamefont
  {Yanagida}},\ }\href {\doibase 10.1016/j.physletb.2009.04.070} {\bibfield
  {journal} {\bibinfo  {journal} {Phys. Lett.}\ }\textbf {\bibinfo {volume}
  {B677}},\ \bibinfo {pages} {59} (\bibinfo {year} {2009})},\ \Eprint
  {http://arxiv.org/abs/0901.2168} {arXiv:0901.2168 [hep-ph]} \BibitemShut
  {NoStop}%
\bibitem [{\citenamefont {Barbier}\ \emph {et~al.}(2005)\citenamefont {Barbier}
  \emph {et~al.}}]{Barbier:2004ez}%
  \BibitemOpen
  \bibfield  {author} {\bibinfo {author} {\bibfnamefont {R.}~\bibnamefont
  {Barbier}} \emph {et~al.},\ }\href {\doibase 10.1016/j.physrep.2005.08.006}
  {\bibfield  {journal} {\bibinfo  {journal} {Phys. Rept.}\ }\textbf {\bibinfo
  {volume} {420}},\ \bibinfo {pages} {1} (\bibinfo {year} {2005})},\ \Eprint
  {http://arxiv.org/abs/hep-ph/0406039} {arXiv:hep-ph/0406039 [hep-ph]}
  \BibitemShut {NoStop}%
\bibitem [{\citenamefont {Christodoulakis}\ and\ \citenamefont
  {Korfiatis}(1991)}]{Christodoulakis:1990zy}%
  \BibitemOpen
  \bibfield  {author} {\bibinfo {author} {\bibfnamefont {T.}~\bibnamefont
  {Christodoulakis}}\ and\ \bibinfo {author} {\bibfnamefont {E.}~\bibnamefont
  {Korfiatis}},\ }\href {\doibase 10.1016/0370-2693(91)91791-S} {\bibfield
  {journal} {\bibinfo  {journal} {Phys. Lett.}\ }\textbf {\bibinfo {volume}
  {B256}},\ \bibinfo {pages} {457} (\bibinfo {year} {1991})}\BibitemShut
  {NoStop}%
\bibitem [{\citenamefont {Fischler}\ \emph {et~al.}(1991)\citenamefont
  {Fischler}, \citenamefont {Giudice}, \citenamefont {Leigh},\ and\
  \citenamefont {Paban}}]{Fischler:1990gn}%
  \BibitemOpen
  \bibfield  {author} {\bibinfo {author} {\bibfnamefont {W.}~\bibnamefont
  {Fischler}}, \bibinfo {author} {\bibfnamefont {G.~F.}\ \bibnamefont
  {Giudice}}, \bibinfo {author} {\bibfnamefont {R.~G.}\ \bibnamefont {Leigh}},
  \ and\ \bibinfo {author} {\bibfnamefont {S.}~\bibnamefont {Paban}},\ }\href
  {\doibase 10.1016/0370-2693(91)91207-C} {\bibfield  {journal} {\bibinfo
  {journal} {Phys. Lett.}\ }\textbf {\bibinfo {volume} {B258}},\ \bibinfo
  {pages} {45} (\bibinfo {year} {1991})}\BibitemShut {NoStop}%
\bibitem [{\citenamefont {Dreiner}\ and\ \citenamefont
  {Ross}(1993)}]{Dreiner:1992vm}%
  \BibitemOpen
  \bibfield  {author} {\bibinfo {author} {\bibfnamefont {H.~K.}\ \bibnamefont
  {Dreiner}}\ and\ \bibinfo {author} {\bibfnamefont {G.~G.}\ \bibnamefont
  {Ross}},\ }\href {\doibase 10.1016/0550-3213(93)90579-E} {\bibfield
  {journal} {\bibinfo  {journal} {Nucl. Phys.}\ }\textbf {\bibinfo {volume}
  {B410}},\ \bibinfo {pages} {188} (\bibinfo {year} {1993})},\ \Eprint
  {http://arxiv.org/abs/hep-ph/9207221} {arXiv:hep-ph/9207221 [hep-ph]}
  \BibitemShut {NoStop}%
\bibitem [{\citenamefont {Endo}\ \emph {et~al.}(2010)\citenamefont {Endo},
  \citenamefont {Hamaguchi},\ and\ \citenamefont {Iwamoto}}]{Endo:2009cv}%
  \BibitemOpen
  \bibfield  {author} {\bibinfo {author} {\bibfnamefont {M.}~\bibnamefont
  {Endo}}, \bibinfo {author} {\bibfnamefont {K.}~\bibnamefont {Hamaguchi}}, \
  and\ \bibinfo {author} {\bibfnamefont {S.}~\bibnamefont {Iwamoto}},\ }\href
  {\doibase 10.1088/1475-7516/2010/02/032} {\bibfield  {journal} {\bibinfo
  {journal} {JCAP}\ }\textbf {\bibinfo {volume} {1002}},\ \bibinfo {pages}
  {032} (\bibinfo {year} {2010})},\ \Eprint {http://arxiv.org/abs/0912.0585}
  {arXiv:0912.0585 [hep-ph]} \BibitemShut {NoStop}%
\bibitem [{\citenamefont {Higaki}\ \emph {et~al.}(2014)\citenamefont {Higaki},
  \citenamefont {Nakayama}, \citenamefont {Saikawa}, \citenamefont
  {Takahashi},\ and\ \citenamefont {Yamaguchi}}]{Higaki:2014eda}%
  \BibitemOpen
  \bibfield  {author} {\bibinfo {author} {\bibfnamefont {T.}~\bibnamefont
  {Higaki}}, \bibinfo {author} {\bibfnamefont {K.}~\bibnamefont {Nakayama}},
  \bibinfo {author} {\bibfnamefont {K.}~\bibnamefont {Saikawa}}, \bibinfo
  {author} {\bibfnamefont {T.}~\bibnamefont {Takahashi}}, \ and\ \bibinfo
  {author} {\bibfnamefont {M.}~\bibnamefont {Yamaguchi}},\ }\href {\doibase
  10.1103/PhysRevD.90.045001} {\bibfield  {journal} {\bibinfo  {journal} {Phys.
  Rev.}\ }\textbf {\bibinfo {volume} {D90}},\ \bibinfo {pages} {045001}
  (\bibinfo {year} {2014})},\ \Eprint {http://arxiv.org/abs/1404.5796}
  {arXiv:1404.5796 [hep-ph]} \BibitemShut {NoStop}%
\bibitem [{\citenamefont {Affleck}\ and\ \citenamefont
  {Dine}(1985)}]{Affleck:1984fy}%
  \BibitemOpen
  \bibfield  {author} {\bibinfo {author} {\bibfnamefont {I.}~\bibnamefont
  {Affleck}}\ and\ \bibinfo {author} {\bibfnamefont {M.}~\bibnamefont {Dine}},\
  }\href {\doibase 10.1016/0550-3213(85)90021-5} {\bibfield  {journal}
  {\bibinfo  {journal} {Nucl. Phys.}\ }\textbf {\bibinfo {volume} {B249}},\
  \bibinfo {pages} {361} (\bibinfo {year} {1985})}\BibitemShut {NoStop}%
\bibitem [{\citenamefont {Dine}\ \emph {et~al.}(1996)\citenamefont {Dine},
  \citenamefont {Randall},\ and\ \citenamefont {Thomas}}]{Dine:1995kz}%
  \BibitemOpen
  \bibfield  {author} {\bibinfo {author} {\bibfnamefont {M.}~\bibnamefont
  {Dine}}, \bibinfo {author} {\bibfnamefont {L.}~\bibnamefont {Randall}}, \
  and\ \bibinfo {author} {\bibfnamefont {S.~D.}\ \bibnamefont {Thomas}},\
  }\href {\doibase 10.1016/0550-3213(95)00538-2} {\bibfield  {journal}
  {\bibinfo  {journal} {Nucl. Phys.}\ }\textbf {\bibinfo {volume} {B458}},\
  \bibinfo {pages} {291} (\bibinfo {year} {1996})},\ \Eprint
  {http://arxiv.org/abs/hep-ph/9507453} {arXiv:hep-ph/9507453 [hep-ph]}
  \BibitemShut {NoStop}%
\bibitem [{\citenamefont {Giusarma}\ \emph {et~al.}(2016)\citenamefont
  {Giusarma}, \citenamefont {Gerbino}, \citenamefont {Mena}, \citenamefont
  {Vagnozzi}, \citenamefont {Ho},\ and\ \citenamefont
  {Freese}}]{Giusarma:2016phn}%
  \BibitemOpen
  \bibfield  {author} {\bibinfo {author} {\bibfnamefont {E.}~\bibnamefont
  {Giusarma}}, \bibinfo {author} {\bibfnamefont {M.}~\bibnamefont {Gerbino}},
  \bibinfo {author} {\bibfnamefont {O.}~\bibnamefont {Mena}}, \bibinfo {author}
  {\bibfnamefont {S.}~\bibnamefont {Vagnozzi}}, \bibinfo {author}
  {\bibfnamefont {S.}~\bibnamefont {Ho}}, \ and\ \bibinfo {author}
  {\bibfnamefont {K.}~\bibnamefont {Freese}},\ }\href@noop {} {\  (\bibinfo
  {year} {2016})},\ \Eprint {http://arxiv.org/abs/1605.04320} {arXiv:1605.04320
  [astro-ph.CO]} \BibitemShut {NoStop}%
\bibitem [{\citenamefont {Ibe}\ \emph {et~al.}(2013)\citenamefont {Ibe},
  \citenamefont {Iwamoto}, \citenamefont {Matsumoto}, \citenamefont {Moroi},\
  and\ \citenamefont {Yokozaki}}]{Ibe:2013nka}%
  \BibitemOpen
  \bibfield  {author} {\bibinfo {author} {\bibfnamefont {M.}~\bibnamefont
  {Ibe}}, \bibinfo {author} {\bibfnamefont {S.}~\bibnamefont {Iwamoto}},
  \bibinfo {author} {\bibfnamefont {S.}~\bibnamefont {Matsumoto}}, \bibinfo
  {author} {\bibfnamefont {T.}~\bibnamefont {Moroi}}, \ and\ \bibinfo {author}
  {\bibfnamefont {N.}~\bibnamefont {Yokozaki}},\ }\href {\doibase
  10.1007/JHEP08(2013)029} {\bibfield  {journal} {\bibinfo  {journal} {JHEP}\
  }\textbf {\bibinfo {volume} {08}},\ \bibinfo {pages} {029} (\bibinfo {year}
  {2013})},\ \Eprint {http://arxiv.org/abs/1304.1483} {arXiv:1304.1483
  [hep-ph]} \BibitemShut {NoStop}%
\bibitem [{\citenamefont {Ishiwata}\ \emph {et~al.}(2008)\citenamefont
  {Ishiwata}, \citenamefont {Matsumoto},\ and\ \citenamefont
  {Moroi}}]{Ishiwata:2008cu}%
  \BibitemOpen
  \bibfield  {author} {\bibinfo {author} {\bibfnamefont {K.}~\bibnamefont
  {Ishiwata}}, \bibinfo {author} {\bibfnamefont {S.}~\bibnamefont {Matsumoto}},
  \ and\ \bibinfo {author} {\bibfnamefont {T.}~\bibnamefont {Moroi}},\ }\href
  {\doibase 10.1103/PhysRevD.78.063505} {\bibfield  {journal} {\bibinfo
  {journal} {Phys. Rev.}\ }\textbf {\bibinfo {volume} {D78}},\ \bibinfo {pages}
  {063505} (\bibinfo {year} {2008})},\ \Eprint {http://arxiv.org/abs/0805.1133}
  {arXiv:0805.1133 [hep-ph]} \BibitemShut {NoStop}%
\bibitem [{\citenamefont {Ishiwata}\ \emph
  {et~al.}(2009{\natexlab{a}})\citenamefont {Ishiwata}, \citenamefont
  {Matsumoto},\ and\ \citenamefont {Moroi}}]{Ishiwata:2008cv}%
  \BibitemOpen
  \bibfield  {author} {\bibinfo {author} {\bibfnamefont {K.}~\bibnamefont
  {Ishiwata}}, \bibinfo {author} {\bibfnamefont {S.}~\bibnamefont {Matsumoto}},
  \ and\ \bibinfo {author} {\bibfnamefont {T.}~\bibnamefont {Moroi}},\ }\href
  {\doibase 10.1016/j.physletb.2009.04.049} {\bibfield  {journal} {\bibinfo
  {journal} {Phys. Lett.}\ }\textbf {\bibinfo {volume} {B675}},\ \bibinfo
  {pages} {446} (\bibinfo {year} {2009}{\natexlab{a}})},\ \Eprint
  {http://arxiv.org/abs/0811.0250} {arXiv:0811.0250 [hep-ph]} \BibitemShut
  {NoStop}%
\bibitem [{\citenamefont {Ishiwata}\ \emph
  {et~al.}(2009{\natexlab{b}})\citenamefont {Ishiwata}, \citenamefont
  {Matsumoto},\ and\ \citenamefont {Moroi}}]{Ishiwata:2009vx}%
  \BibitemOpen
  \bibfield  {author} {\bibinfo {author} {\bibfnamefont {K.}~\bibnamefont
  {Ishiwata}}, \bibinfo {author} {\bibfnamefont {S.}~\bibnamefont {Matsumoto}},
  \ and\ \bibinfo {author} {\bibfnamefont {T.}~\bibnamefont {Moroi}},\ }\href
  {\doibase 10.1088/1126-6708/2009/05/110} {\bibfield  {journal} {\bibinfo
  {journal} {JHEP}\ }\textbf {\bibinfo {volume} {05}},\ \bibinfo {pages} {110}
  (\bibinfo {year} {2009}{\natexlab{b}})},\ \Eprint
  {http://arxiv.org/abs/0903.0242} {arXiv:0903.0242 [hep-ph]} \BibitemShut
  {NoStop}%
\bibitem [{\citenamefont {Delahaye}\ and\ \citenamefont
  {Grefe}(2013)}]{Delahaye:2013yqa}%
  \BibitemOpen
  \bibfield  {author} {\bibinfo {author} {\bibfnamefont {T.}~\bibnamefont
  {Delahaye}}\ and\ \bibinfo {author} {\bibfnamefont {M.}~\bibnamefont
  {Grefe}},\ }\href {\doibase 10.1088/1475-7516/2013/12/045} {\bibfield
  {journal} {\bibinfo  {journal} {JCAP}\ }\textbf {\bibinfo {volume} {1312}},\
  \bibinfo {pages} {045} (\bibinfo {year} {2013})},\ \Eprint
  {http://arxiv.org/abs/1305.7183} {arXiv:1305.7183 [hep-ph]} \BibitemShut
  {NoStop}%
\bibitem [{\citenamefont {Ibarra}\ and\ \citenamefont
  {Tran}(2008)}]{Ibarra:2007wg}%
  \BibitemOpen
  \bibfield  {author} {\bibinfo {author} {\bibfnamefont {A.}~\bibnamefont
  {Ibarra}}\ and\ \bibinfo {author} {\bibfnamefont {D.}~\bibnamefont {Tran}},\
  }\href {\doibase 10.1103/PhysRevLett.100.061301} {\bibfield  {journal}
  {\bibinfo  {journal} {Phys. Rev. Lett.}\ }\textbf {\bibinfo {volume} {100}},\
  \bibinfo {pages} {061301} (\bibinfo {year} {2008})},\ \Eprint
  {http://arxiv.org/abs/0709.4593} {arXiv:0709.4593 [astro-ph]} \BibitemShut
  {NoStop}%
\bibitem [{\citenamefont {Ishiwata}\ \emph
  {et~al.}(2009{\natexlab{c}})\citenamefont {Ishiwata}, \citenamefont
  {Matsumoto},\ and\ \citenamefont {Moroi}}]{Ishiwata:2009dk}%
  \BibitemOpen
  \bibfield  {author} {\bibinfo {author} {\bibfnamefont {K.}~\bibnamefont
  {Ishiwata}}, \bibinfo {author} {\bibfnamefont {S.}~\bibnamefont {Matsumoto}},
  \ and\ \bibinfo {author} {\bibfnamefont {T.}~\bibnamefont {Moroi}},\ }\href
  {\doibase 10.1016/j.physletb.2009.07.004} {\bibfield  {journal} {\bibinfo
  {journal} {Phys. Lett.}\ }\textbf {\bibinfo {volume} {B679}},\ \bibinfo
  {pages} {1} (\bibinfo {year} {2009}{\natexlab{c}})},\ \Eprint
  {http://arxiv.org/abs/0905.4593} {arXiv:0905.4593 [astro-ph.CO]} \BibitemShut
  {NoStop}%
\bibitem [{\citenamefont {Carquin}\ \emph {et~al.}(2016)\citenamefont
  {Carquin}, \citenamefont {Diaz}, \citenamefont {Gomez-Vargas}, \citenamefont
  {Panes},\ and\ \citenamefont {Viaux}}]{Carquin:2015uma}%
  \BibitemOpen
  \bibfield  {author} {\bibinfo {author} {\bibfnamefont {E.}~\bibnamefont
  {Carquin}}, \bibinfo {author} {\bibfnamefont {M.~A.}\ \bibnamefont {Diaz}},
  \bibinfo {author} {\bibfnamefont {G.~A.}\ \bibnamefont {Gomez-Vargas}},
  \bibinfo {author} {\bibfnamefont {B.}~\bibnamefont {Panes}}, \ and\ \bibinfo
  {author} {\bibfnamefont {N.}~\bibnamefont {Viaux}},\ }\href {\doibase
  10.1016/j.dark.2015.10.002} {\bibfield  {journal} {\bibinfo  {journal} {Phys.
  Dark Univ.}\ }\textbf {\bibinfo {volume} {11}},\ \bibinfo {pages} {1}
  (\bibinfo {year} {2016})},\ \Eprint {http://arxiv.org/abs/1501.05932}
  {arXiv:1501.05932 [hep-ph]} \BibitemShut {NoStop}%
\bibitem [{\citenamefont {Ando}\ and\ \citenamefont
  {Ishiwata}(2015)}]{Ando:2015qda}%
  \BibitemOpen
  \bibfield  {author} {\bibinfo {author} {\bibfnamefont {S.}~\bibnamefont
  {Ando}}\ and\ \bibinfo {author} {\bibfnamefont {K.}~\bibnamefont
  {Ishiwata}},\ }\href {\doibase 10.1088/1475-7516/2015/05/024} {\bibfield
  {journal} {\bibinfo  {journal} {JCAP}\ }\textbf {\bibinfo {volume} {1505}},\
  \bibinfo {pages} {024} (\bibinfo {year} {2015})},\ \Eprint
  {http://arxiv.org/abs/1502.02007} {arXiv:1502.02007 [astro-ph.CO]}
  \BibitemShut {NoStop}%
\bibitem [{\citenamefont {Ando}\ and\ \citenamefont
  {Ishiwata}(2016)}]{Ando:2016ang}%
  \BibitemOpen
  \bibfield  {author} {\bibinfo {author} {\bibfnamefont {S.}~\bibnamefont
  {Ando}}\ and\ \bibinfo {author} {\bibfnamefont {K.}~\bibnamefont
  {Ishiwata}},\ }\href {\doibase 10.1088/1475-7516/2016/06/045} {\bibfield
  {journal} {\bibinfo  {journal} {JCAP}\ }\textbf {\bibinfo {volume} {1606}},\
  \bibinfo {pages} {045} (\bibinfo {year} {2016})},\ \Eprint
  {http://arxiv.org/abs/1604.02263} {arXiv:1604.02263 [hep-ph]} \BibitemShut
  {NoStop}%
\bibitem [{\citenamefont {Covi}\ \emph {et~al.}(2009)\citenamefont {Covi},
  \citenamefont {Grefe}, \citenamefont {Ibarra},\ and\ \citenamefont
  {Tran}}]{Covi:2008jy}%
  \BibitemOpen
  \bibfield  {author} {\bibinfo {author} {\bibfnamefont {L.}~\bibnamefont
  {Covi}}, \bibinfo {author} {\bibfnamefont {M.}~\bibnamefont {Grefe}},
  \bibinfo {author} {\bibfnamefont {A.}~\bibnamefont {Ibarra}}, \ and\ \bibinfo
  {author} {\bibfnamefont {D.}~\bibnamefont {Tran}},\ }\href {\doibase
  10.1088/1475-7516/2009/01/029} {\bibfield  {journal} {\bibinfo  {journal}
  {JCAP}\ }\textbf {\bibinfo {volume} {0901}},\ \bibinfo {pages} {029}
  (\bibinfo {year} {2009})},\ \Eprint {http://arxiv.org/abs/0809.5030}
  {arXiv:0809.5030 [hep-ph]} \BibitemShut {NoStop}%
\bibitem [{\citenamefont {Ackermann}\ \emph
  {et~al.}(2015{\natexlab{a}})\citenamefont {Ackermann} \emph
  {et~al.}}]{Ackermann:2014usa}%
  \BibitemOpen
  \bibfield  {author} {\bibinfo {author} {\bibfnamefont {M.}~\bibnamefont
  {Ackermann}} \emph {et~al.} (\bibinfo {collaboration} {Fermi-LAT}),\ }\href
  {\doibase 10.1088/0004-637X/799/1/86} {\bibfield  {journal} {\bibinfo
  {journal} {Astrophys. J.}\ }\textbf {\bibinfo {volume} {799}},\ \bibinfo
  {pages} {86} (\bibinfo {year} {2015}{\natexlab{a}})},\ \Eprint
  {http://arxiv.org/abs/1410.3696} {arXiv:1410.3696 [astro-ph.HE]} \BibitemShut
  {NoStop}%
\bibitem [{\citenamefont {Rychkov}\ and\ \citenamefont
  {Strumia}(2007)}]{Rychkov:2007uq}%
  \BibitemOpen
  \bibfield  {author} {\bibinfo {author} {\bibfnamefont {V.~S.}\ \bibnamefont
  {Rychkov}}\ and\ \bibinfo {author} {\bibfnamefont {A.}~\bibnamefont
  {Strumia}},\ }\href {\doibase 10.1103/PhysRevD.75.075011} {\bibfield
  {journal} {\bibinfo  {journal} {Phys. Rev.}\ }\textbf {\bibinfo {volume}
  {D75}},\ \bibinfo {pages} {075011} (\bibinfo {year} {2007})},\ \Eprint
  {http://arxiv.org/abs/hep-ph/0701104} {arXiv:hep-ph/0701104 [hep-ph]}
  \BibitemShut {NoStop}%
\bibitem [{\citenamefont {Ellis}\ \emph {et~al.}(2016)\citenamefont {Ellis},
  \citenamefont {Garcia}, \citenamefont {Nanopoulos}, \citenamefont {Olive},\
  and\ \citenamefont {Peloso}}]{Ellis:2015jpg}%
  \BibitemOpen
  \bibfield  {author} {\bibinfo {author} {\bibfnamefont {J.}~\bibnamefont
  {Ellis}}, \bibinfo {author} {\bibfnamefont {M.~A.~G.}\ \bibnamefont
  {Garcia}}, \bibinfo {author} {\bibfnamefont {D.~V.}\ \bibnamefont
  {Nanopoulos}}, \bibinfo {author} {\bibfnamefont {K.~A.}\ \bibnamefont
  {Olive}}, \ and\ \bibinfo {author} {\bibfnamefont {M.}~\bibnamefont
  {Peloso}},\ }\href {\doibase 10.1088/1475-7516/2016/03/008} {\bibfield
  {journal} {\bibinfo  {journal} {JCAP}\ }\textbf {\bibinfo {volume} {1603}},\
  \bibinfo {pages} {008} (\bibinfo {year} {2016})},\ \Eprint
  {http://arxiv.org/abs/1512.05701} {arXiv:1512.05701 [astro-ph.CO]}
  \BibitemShut {NoStop}%
\bibitem [{\citenamefont {Pradler}\ and\ \citenamefont
  {Steffen}(2007)}]{Pradler:2006qh}%
  \BibitemOpen
  \bibfield  {author} {\bibinfo {author} {\bibfnamefont {J.}~\bibnamefont
  {Pradler}}\ and\ \bibinfo {author} {\bibfnamefont {F.~D.}\ \bibnamefont
  {Steffen}},\ }\href {\doibase 10.1103/PhysRevD.75.023509} {\bibfield
  {journal} {\bibinfo  {journal} {Phys. Rev.}\ }\textbf {\bibinfo {volume}
  {D75}},\ \bibinfo {pages} {023509} (\bibinfo {year} {2007})},\ \Eprint
  {http://arxiv.org/abs/hep-ph/0608344} {arXiv:hep-ph/0608344 [hep-ph]}
  \BibitemShut {NoStop}%
\bibitem [{\citenamefont {Allanach}(2002)}]{Allanach:2001kg}%
  \BibitemOpen
  \bibfield  {author} {\bibinfo {author} {\bibfnamefont {B.~C.}\ \bibnamefont
  {Allanach}},\ }\href {\doibase 10.1016/S0010-4655(01)00460-X} {\bibfield
  {journal} {\bibinfo  {journal} {Comput. Phys. Commun.}\ }\textbf {\bibinfo
  {volume} {143}},\ \bibinfo {pages} {305} (\bibinfo {year} {2002})},\ \Eprint
  {http://arxiv.org/abs/hep-ph/0104145} {arXiv:hep-ph/0104145 [hep-ph]}
  \BibitemShut {NoStop}%
\bibitem [{\citenamefont {Antusch}\ and\ \citenamefont
  {Teixeira}(2007)}]{Antusch:2006gy}%
  \BibitemOpen
  \bibfield  {author} {\bibinfo {author} {\bibfnamefont {S.}~\bibnamefont
  {Antusch}}\ and\ \bibinfo {author} {\bibfnamefont {A.~M.}\ \bibnamefont
  {Teixeira}},\ }\href {\doibase 10.1088/1475-7516/2007/02/024} {\bibfield
  {journal} {\bibinfo  {journal} {JCAP}\ }\textbf {\bibinfo {volume} {0702}},\
  \bibinfo {pages} {024} (\bibinfo {year} {2007})},\ \Eprint
  {http://arxiv.org/abs/hep-ph/0611232} {arXiv:hep-ph/0611232 [hep-ph]}
  \BibitemShut {NoStop}%
\bibitem [{\citenamefont {Ade}\ \emph {et~al.}(2015)\citenamefont {Ade} \emph
  {et~al.}}]{Ade:2015lrj}%
  \BibitemOpen
  \bibfield  {author} {\bibinfo {author} {\bibfnamefont {P.~A.~R.}\
  \bibnamefont {Ade}} \emph {et~al.} (\bibinfo {collaboration} {Planck}),\
  }\href@noop {} {\  (\bibinfo {year} {2015})},\ \Eprint
  {http://arxiv.org/abs/1502.02114} {arXiv:1502.02114 [astro-ph.CO]}
  \BibitemShut {NoStop}%
\bibitem [{\citenamefont {Feng}\ \emph
  {et~al.}(2003{\natexlab{a}})\citenamefont {Feng}, \citenamefont {Rajaraman},\
  and\ \citenamefont {Takayama}}]{Feng:2003xh}%
  \BibitemOpen
  \bibfield  {author} {\bibinfo {author} {\bibfnamefont {J.~L.}\ \bibnamefont
  {Feng}}, \bibinfo {author} {\bibfnamefont {A.}~\bibnamefont {Rajaraman}}, \
  and\ \bibinfo {author} {\bibfnamefont {F.}~\bibnamefont {Takayama}},\ }\href
  {\doibase 10.1103/PhysRevLett.91.011302} {\bibfield  {journal} {\bibinfo
  {journal} {Phys. Rev. Lett.}\ }\textbf {\bibinfo {volume} {91}},\ \bibinfo
  {pages} {011302} (\bibinfo {year} {2003}{\natexlab{a}})},\ \Eprint
  {http://arxiv.org/abs/hep-ph/0302215} {arXiv:hep-ph/0302215 [hep-ph]}
  \BibitemShut {NoStop}%
\bibitem [{\citenamefont {Feng}\ \emph
  {et~al.}(2003{\natexlab{b}})\citenamefont {Feng}, \citenamefont {Rajaraman},\
  and\ \citenamefont {Takayama}}]{Feng:2003uy}%
  \BibitemOpen
  \bibfield  {author} {\bibinfo {author} {\bibfnamefont {J.~L.}\ \bibnamefont
  {Feng}}, \bibinfo {author} {\bibfnamefont {A.}~\bibnamefont {Rajaraman}}, \
  and\ \bibinfo {author} {\bibfnamefont {F.}~\bibnamefont {Takayama}},\ }\href
  {\doibase 10.1103/PhysRevD.68.063504} {\bibfield  {journal} {\bibinfo
  {journal} {Phys. Rev.}\ }\textbf {\bibinfo {volume} {D68}},\ \bibinfo {pages}
  {063504} (\bibinfo {year} {2003}{\natexlab{b}})},\ \Eprint
  {http://arxiv.org/abs/hep-ph/0306024} {arXiv:hep-ph/0306024 [hep-ph]}
  \BibitemShut {NoStop}%
\bibitem [{\citenamefont {Fujii}\ \emph {et~al.}(2004)\citenamefont {Fujii},
  \citenamefont {Ibe},\ and\ \citenamefont {Yanagida}}]{Fujii:2003nr}%
  \BibitemOpen
  \bibfield  {author} {\bibinfo {author} {\bibfnamefont {M.}~\bibnamefont
  {Fujii}}, \bibinfo {author} {\bibfnamefont {M.}~\bibnamefont {Ibe}}, \ and\
  \bibinfo {author} {\bibfnamefont {T.}~\bibnamefont {Yanagida}},\ }\href
  {\doibase 10.1016/j.physletb.2003.10.092} {\bibfield  {journal} {\bibinfo
  {journal} {Phys. Lett.}\ }\textbf {\bibinfo {volume} {B579}},\ \bibinfo
  {pages} {6} (\bibinfo {year} {2004})},\ \Eprint
  {http://arxiv.org/abs/hep-ph/0310142} {arXiv:hep-ph/0310142 [hep-ph]}
  \BibitemShut {NoStop}%
\bibitem [{\citenamefont {Feng}\ \emph
  {et~al.}(2004{\natexlab{a}})\citenamefont {Feng}, \citenamefont {Su},\ and\
  \citenamefont {Takayama}}]{Feng:2004zu}%
  \BibitemOpen
  \bibfield  {author} {\bibinfo {author} {\bibfnamefont {J.~L.}\ \bibnamefont
  {Feng}}, \bibinfo {author} {\bibfnamefont {S.-f.}\ \bibnamefont {Su}}, \ and\
  \bibinfo {author} {\bibfnamefont {F.}~\bibnamefont {Takayama}},\ }\href
  {\doibase 10.1103/PhysRevD.70.063514} {\bibfield  {journal} {\bibinfo
  {journal} {Phys. Rev.}\ }\textbf {\bibinfo {volume} {D70}},\ \bibinfo {pages}
  {063514} (\bibinfo {year} {2004}{\natexlab{a}})},\ \Eprint
  {http://arxiv.org/abs/hep-ph/0404198} {arXiv:hep-ph/0404198 [hep-ph]}
  \BibitemShut {NoStop}%
\bibitem [{\citenamefont {Feng}\ \emph
  {et~al.}(2004{\natexlab{b}})\citenamefont {Feng}, \citenamefont {Su},\ and\
  \citenamefont {Takayama}}]{Feng:2004mt}%
  \BibitemOpen
  \bibfield  {author} {\bibinfo {author} {\bibfnamefont {J.~L.}\ \bibnamefont
  {Feng}}, \bibinfo {author} {\bibfnamefont {S.}~\bibnamefont {Su}}, \ and\
  \bibinfo {author} {\bibfnamefont {F.}~\bibnamefont {Takayama}},\ }\href
  {\doibase 10.1103/PhysRevD.70.075019} {\bibfield  {journal} {\bibinfo
  {journal} {Phys. Rev.}\ }\textbf {\bibinfo {volume} {D70}},\ \bibinfo {pages}
  {075019} (\bibinfo {year} {2004}{\natexlab{b}})},\ \Eprint
  {http://arxiv.org/abs/hep-ph/0404231} {arXiv:hep-ph/0404231 [hep-ph]}
  \BibitemShut {NoStop}%
\bibitem [{\citenamefont {Heisig}(2014)}]{Heisig:2013sva}%
  \BibitemOpen
  \bibfield  {author} {\bibinfo {author} {\bibfnamefont {J.}~\bibnamefont
  {Heisig}},\ }\href {\doibase 10.1088/1475-7516/2014/04/023} {\bibfield
  {journal} {\bibinfo  {journal} {JCAP}\ }\textbf {\bibinfo {volume} {1404}},\
  \bibinfo {pages} {023} (\bibinfo {year} {2014})},\ \Eprint
  {http://arxiv.org/abs/1310.6352} {arXiv:1310.6352 [hep-ph]} \BibitemShut
  {NoStop}%
\bibitem [{\citenamefont {Arbey}\ \emph {et~al.}(2015)\citenamefont {Arbey},
  \citenamefont {Battaglia}, \citenamefont {Covi}, \citenamefont {Hasenkamp},\
  and\ \citenamefont {Mahmoudi}}]{Arvey:2015nra}%
  \BibitemOpen
  \bibfield  {author} {\bibinfo {author} {\bibfnamefont {A.}~\bibnamefont
  {Arbey}}, \bibinfo {author} {\bibfnamefont {M.}~\bibnamefont {Battaglia}},
  \bibinfo {author} {\bibfnamefont {L.}~\bibnamefont {Covi}}, \bibinfo {author}
  {\bibfnamefont {J.}~\bibnamefont {Hasenkamp}}, \ and\ \bibinfo {author}
  {\bibfnamefont {F.}~\bibnamefont {Mahmoudi}},\ }\href {\doibase
  10.1103/PhysRevD.92.115008} {\bibfield  {journal} {\bibinfo  {journal} {Phys.
  Rev.}\ }\textbf {\bibinfo {volume} {D92}},\ \bibinfo {pages} {115008}
  (\bibinfo {year} {2015})},\ \Eprint {http://arxiv.org/abs/1505.04595}
  {arXiv:1505.04595 [hep-ph]} \BibitemShut {NoStop}%
\bibitem [{\citenamefont {Kawasaki}\ \emph {et~al.}(2005)\citenamefont
  {Kawasaki}, \citenamefont {Kohri},\ and\ \citenamefont
  {Moroi}}]{Kawasaki:2004qu}%
  \BibitemOpen
  \bibfield  {author} {\bibinfo {author} {\bibfnamefont {M.}~\bibnamefont
  {Kawasaki}}, \bibinfo {author} {\bibfnamefont {K.}~\bibnamefont {Kohri}}, \
  and\ \bibinfo {author} {\bibfnamefont {T.}~\bibnamefont {Moroi}},\ }\href
  {\doibase 10.1103/PhysRevD.71.083502} {\bibfield  {journal} {\bibinfo
  {journal} {Phys. Rev.}\ }\textbf {\bibinfo {volume} {D71}},\ \bibinfo {pages}
  {083502} (\bibinfo {year} {2005})},\ \Eprint
  {http://arxiv.org/abs/astro-ph/0408426} {arXiv:astro-ph/0408426 [astro-ph]}
  \BibitemShut {NoStop}%
\bibitem [{\citenamefont {Jedamzik}(2006)}]{Jedamzik:2006xz}%
  \BibitemOpen
  \bibfield  {author} {\bibinfo {author} {\bibfnamefont {K.}~\bibnamefont
  {Jedamzik}},\ }\href {\doibase 10.1103/PhysRevD.74.103509} {\bibfield
  {journal} {\bibinfo  {journal} {Phys. Rev.}\ }\textbf {\bibinfo {volume}
  {D74}},\ \bibinfo {pages} {103509} (\bibinfo {year} {2006})},\ \Eprint
  {http://arxiv.org/abs/hep-ph/0604251} {arXiv:hep-ph/0604251 [hep-ph]}
  \BibitemShut {NoStop}%
\bibitem [{\citenamefont {Hirsch}\ \emph {et~al.}(2005)\citenamefont {Hirsch},
  \citenamefont {Porod},\ and\ \citenamefont {Restrepo}}]{Hirsch:2005ag}%
  \BibitemOpen
  \bibfield  {author} {\bibinfo {author} {\bibfnamefont {M.}~\bibnamefont
  {Hirsch}}, \bibinfo {author} {\bibfnamefont {W.}~\bibnamefont {Porod}}, \
  and\ \bibinfo {author} {\bibfnamefont {D.}~\bibnamefont {Restrepo}},\ }\href
  {\doibase 10.1088/1126-6708/2005/03/062} {\bibfield  {journal} {\bibinfo
  {journal} {JHEP}\ }\textbf {\bibinfo {volume} {03}},\ \bibinfo {pages} {062}
  (\bibinfo {year} {2005})},\ \Eprint {http://arxiv.org/abs/hep-ph/0503059}
  {arXiv:hep-ph/0503059 [hep-ph]} \BibitemShut {NoStop}%
\bibitem [{ATL(2016)}]{ATLAS:2016kts}%
  \BibitemOpen
  \href@noop {} {\bibfield  {journal} {\bibinfo  {journal} {ATLAS
  collaboration, ATLAS-CONF-2016-078}\ } (\bibinfo {year} {2016})}\BibitemShut
  {NoStop}%
\bibitem [{\citenamefont {Arbey}\ \emph {et~al.}(2016)\citenamefont {Arbey},
  \citenamefont {Battaglia},\ and\ \citenamefont {Mahmoudi}}]{Arbey:2015hca}%
  \BibitemOpen
  \bibfield  {author} {\bibinfo {author} {\bibfnamefont {A.}~\bibnamefont
  {Arbey}}, \bibinfo {author} {\bibfnamefont {M.}~\bibnamefont {Battaglia}}, \
  and\ \bibinfo {author} {\bibfnamefont {F.}~\bibnamefont {Mahmoudi}},\ }\href
  {\doibase 10.1103/PhysRevD.94.055015} {\bibfield  {journal} {\bibinfo
  {journal} {Phys. Rev.}\ }\textbf {\bibinfo {volume} {D94}},\ \bibinfo {pages}
  {055015} (\bibinfo {year} {2016})},\ \Eprint
  {http://arxiv.org/abs/1506.02148} {arXiv:1506.02148 [hep-ph]} \BibitemShut
  {NoStop}%
\bibitem [{CMS(2015)}]{CMS:2015kdx}%
  \BibitemOpen
  \href@noop {} {\bibfield  {journal} {\bibinfo  {journal} {CMS Collaboration,
  CMS-PAS-EXO-15-010}\ } (\bibinfo {year} {2015})}\BibitemShut {NoStop}%
\bibitem [{\citenamefont {Borschensky}\ \emph {et~al.}(2014)\citenamefont
  {Borschensky}, \citenamefont {Kr{\"a}mer}, \citenamefont {Kulesza},
  \citenamefont {Mangano}, \citenamefont {Padhi}, \citenamefont {Plehn},\ and\
  \citenamefont {Portell}}]{Borschensky:2014cia}%
  \BibitemOpen
  \bibfield  {author} {\bibinfo {author} {\bibfnamefont {C.}~\bibnamefont
  {Borschensky}}, \bibinfo {author} {\bibfnamefont {M.}~\bibnamefont
  {Kr{\"a}mer}}, \bibinfo {author} {\bibfnamefont {A.}~\bibnamefont {Kulesza}},
  \bibinfo {author} {\bibfnamefont {M.}~\bibnamefont {Mangano}}, \bibinfo
  {author} {\bibfnamefont {S.}~\bibnamefont {Padhi}}, \bibinfo {author}
  {\bibfnamefont {T.}~\bibnamefont {Plehn}}, \ and\ \bibinfo {author}
  {\bibfnamefont {X.}~\bibnamefont {Portell}},\ }\href {\doibase
  10.1140/epjc/s10052-014-3174-y} {\bibfield  {journal} {\bibinfo  {journal}
  {Eur. Phys. J.}\ }\textbf {\bibinfo {volume} {C74}},\ \bibinfo {pages} {3174}
  (\bibinfo {year} {2014})},\ \Eprint {http://arxiv.org/abs/1407.5066}
  {arXiv:1407.5066 [hep-ph]} \BibitemShut {NoStop}%
\bibitem [{CMS(2016)}]{CMS:2016mwj}%
  \BibitemOpen
  \href@noop {} {\bibfield  {journal} {\bibinfo  {journal} {CMS Collaboration,
  CMS-PAS-SUS-16-014}\ } (\bibinfo {year} {2016})}\BibitemShut {NoStop}%
\bibitem [{\citenamefont {Cohen}\ \emph {et~al.}(2013)\citenamefont {Cohen},
  \citenamefont {Golling}, \citenamefont {Hance}, \citenamefont {Henrichs},
  \citenamefont {Howe}, \citenamefont {Loyal}, \citenamefont {Padhi},\ and\
  \citenamefont {Wacker}}]{Cohen:2013zla}%
  \BibitemOpen
  \bibfield  {author} {\bibinfo {author} {\bibfnamefont {T.}~\bibnamefont
  {Cohen}}, \bibinfo {author} {\bibfnamefont {T.}~\bibnamefont {Golling}},
  \bibinfo {author} {\bibfnamefont {M.}~\bibnamefont {Hance}}, \bibinfo
  {author} {\bibfnamefont {A.}~\bibnamefont {Henrichs}}, \bibinfo {author}
  {\bibfnamefont {K.}~\bibnamefont {Howe}}, \bibinfo {author} {\bibfnamefont
  {J.}~\bibnamefont {Loyal}}, \bibinfo {author} {\bibfnamefont
  {S.}~\bibnamefont {Padhi}}, \ and\ \bibinfo {author} {\bibfnamefont {J.~G.}\
  \bibnamefont {Wacker}},\ }in\ \href
  {http://inspirehep.net/record/1256319/files/arXiv:1310.0077.pdf} {\emph
  {\bibinfo {booktitle} {{Proceedings, Community Summer Study 2013: Snowmass on
  the Mississippi (CSS2013): Minneapolis, MN, USA, July 29-August 6, 2013}}}}\
  (\bibinfo {year} {2013})\ \Eprint {http://arxiv.org/abs/1310.0077}
  {arXiv:1310.0077 [hep-ph]} \BibitemShut {NoStop}%
\bibitem [{\citenamefont {Farrar}(1998)}]{Farrar:1997ns}%
  \BibitemOpen
  \bibfield  {author} {\bibinfo {author} {\bibfnamefont {G.~R.}\ \bibnamefont
  {Farrar}},\ }\bibfield  {booktitle} {\emph {\bibinfo {booktitle}
  {{Supersymmetry in physics. Proceedings, 5th International Conference,
  SUSY'97, Philadelphia, USA, May 27-31, 1997}}},\ }\href {\doibase
  10.1016/S0920-5632(97)00689-0} {\bibfield  {journal} {\bibinfo  {journal}
  {Nucl. Phys. Proc. Suppl.}\ }\textbf {\bibinfo {volume} {62}},\ \bibinfo
  {pages} {485} (\bibinfo {year} {1998})},\ \Eprint
  {http://arxiv.org/abs/hep-ph/9710277} {arXiv:hep-ph/9710277 [hep-ph]}
  \BibitemShut {NoStop}%
\bibitem [{\citenamefont {Kraan}\ \emph {et~al.}(2007)\citenamefont {Kraan},
  \citenamefont {Hansen},\ and\ \citenamefont {Nevski}}]{Kraan:2005ji}%
  \BibitemOpen
  \bibfield  {author} {\bibinfo {author} {\bibfnamefont {A.~C.}\ \bibnamefont
  {Kraan}}, \bibinfo {author} {\bibfnamefont {J.~B.}\ \bibnamefont {Hansen}}, \
  and\ \bibinfo {author} {\bibfnamefont {P.}~\bibnamefont {Nevski}},\ }\href
  {\doibase 10.1140/epjc/s10052-006-0162-x} {\bibfield  {journal} {\bibinfo
  {journal} {Eur. Phys. J.}\ }\textbf {\bibinfo {volume} {C49}},\ \bibinfo
  {pages} {623} (\bibinfo {year} {2007})},\ \Eprint
  {http://arxiv.org/abs/hep-ex/0511014} {arXiv:hep-ex/0511014 [hep-ex]}
  \BibitemShut {NoStop}%
\bibitem [{\citenamefont {Ackermann}\ \emph
  {et~al.}(2015{\natexlab{b}})\citenamefont {Ackermann} \emph
  {et~al.}}]{Ackermann:2015lka}%
  \BibitemOpen
  \bibfield  {author} {\bibinfo {author} {\bibfnamefont {M.}~\bibnamefont
  {Ackermann}} \emph {et~al.} (\bibinfo {collaboration} {Fermi-LAT}),\ }\href
  {\doibase 10.1103/PhysRevD.91.122002} {\bibfield  {journal} {\bibinfo
  {journal} {Phys. Rev.}\ }\textbf {\bibinfo {volume} {D91}},\ \bibinfo {pages}
  {122002} (\bibinfo {year} {2015}{\natexlab{b}})},\ \Eprint
  {http://arxiv.org/abs/1506.00013} {arXiv:1506.00013 [astro-ph.HE]}
  \BibitemShut {NoStop}%
\bibitem [{\citenamefont {Asai}\ \emph {et~al.}(2011)\citenamefont {Asai},
  \citenamefont {Azuma}, \citenamefont {Endo}, \citenamefont {Hamaguchi},\ and\
  \citenamefont {Iwamoto}}]{Asai:2011wy}%
  \BibitemOpen
  \bibfield  {author} {\bibinfo {author} {\bibfnamefont {S.}~\bibnamefont
  {Asai}}, \bibinfo {author} {\bibfnamefont {Y.}~\bibnamefont {Azuma}},
  \bibinfo {author} {\bibfnamefont {M.}~\bibnamefont {Endo}}, \bibinfo {author}
  {\bibfnamefont {K.}~\bibnamefont {Hamaguchi}}, \ and\ \bibinfo {author}
  {\bibfnamefont {S.}~\bibnamefont {Iwamoto}},\ }\href {\doibase
  10.1007/JHEP12(2011)041} {\bibfield  {journal} {\bibinfo  {journal} {JHEP}\
  }\textbf {\bibinfo {volume} {12}},\ \bibinfo {pages} {041} (\bibinfo {year}
  {2011})},\ \Eprint {http://arxiv.org/abs/1103.1881} {arXiv:1103.1881
  [hep-ph]} \BibitemShut {NoStop}%
\bibitem [{\citenamefont {Graham}\ \emph {et~al.}(2012)\citenamefont {Graham},
  \citenamefont {Kaplan}, \citenamefont {Rajendran},\ and\ \citenamefont
  {Saraswat}}]{Graham:2012th}%
  \BibitemOpen
  \bibfield  {author} {\bibinfo {author} {\bibfnamefont {P.~W.}\ \bibnamefont
  {Graham}}, \bibinfo {author} {\bibfnamefont {D.~E.}\ \bibnamefont {Kaplan}},
  \bibinfo {author} {\bibfnamefont {S.}~\bibnamefont {Rajendran}}, \ and\
  \bibinfo {author} {\bibfnamefont {P.}~\bibnamefont {Saraswat}},\ }\href
  {\doibase 10.1007/JHEP07(2012)149} {\bibfield  {journal} {\bibinfo  {journal}
  {JHEP}\ }\textbf {\bibinfo {volume} {07}},\ \bibinfo {pages} {149} (\bibinfo
  {year} {2012})},\ \Eprint {http://arxiv.org/abs/1204.6038} {arXiv:1204.6038
  [hep-ph]} \BibitemShut {NoStop}%
\bibitem [{\citenamefont {Csaki}\ \emph {et~al.}(2015)\citenamefont {Csaki},
  \citenamefont {Kuflik}, \citenamefont {Lombardo}, \citenamefont {Slone},\
  and\ \citenamefont {Volansky}}]{Csaki:2015uza}%
  \BibitemOpen
  \bibfield  {author} {\bibinfo {author} {\bibfnamefont {C.}~\bibnamefont
  {Csaki}}, \bibinfo {author} {\bibfnamefont {E.}~\bibnamefont {Kuflik}},
  \bibinfo {author} {\bibfnamefont {S.}~\bibnamefont {Lombardo}}, \bibinfo
  {author} {\bibfnamefont {O.}~\bibnamefont {Slone}}, \ and\ \bibinfo {author}
  {\bibfnamefont {T.}~\bibnamefont {Volansky}},\ }\href {\doibase
  10.1007/JHEP08(2015)016} {\bibfield  {journal} {\bibinfo  {journal} {JHEP}\
  }\textbf {\bibinfo {volume} {08}},\ \bibinfo {pages} {016} (\bibinfo {year}
  {2015})},\ \Eprint {http://arxiv.org/abs/1505.00784} {arXiv:1505.00784
  [hep-ph]} \BibitemShut {NoStop}%
\bibitem [{The(2013)}]{TheATLAScollaboration:2013yia}%
  \BibitemOpen
  \href@noop {} {\bibfield  {journal} {\bibinfo  {journal} {The ATLAS
  collaboration, ATLAS-CONF-2013-092}\ } (\bibinfo {year} {2013})}\BibitemShut
  {NoStop}%
\bibitem [{\citenamefont {Aad}\ \emph {et~al.}(2015)\citenamefont {Aad} \emph
  {et~al.}}]{Aad:2015rba}%
  \BibitemOpen
  \bibfield  {author} {\bibinfo {author} {\bibfnamefont {G.}~\bibnamefont
  {Aad}} \emph {et~al.} (\bibinfo {collaboration} {ATLAS}),\ }\href {\doibase
  10.1103/PhysRevD.92.072004} {\bibfield  {journal} {\bibinfo  {journal} {Phys.
  Rev.}\ }\textbf {\bibinfo {volume} {D92}},\ \bibinfo {pages} {072004}
  (\bibinfo {year} {2015})},\ \Eprint {http://arxiv.org/abs/1504.05162}
  {arXiv:1504.05162 [hep-ex]} \BibitemShut {NoStop}%
\bibitem [{\citenamefont {Evans}\ and\ \citenamefont
  {Shelton}(2016)}]{Evans:2016zau}%
  \BibitemOpen
  \bibfield  {author} {\bibinfo {author} {\bibfnamefont {J.~A.}\ \bibnamefont
  {Evans}}\ and\ \bibinfo {author} {\bibfnamefont {J.}~\bibnamefont
  {Shelton}},\ }\href {\doibase 10.1007/JHEP04(2016)056} {\bibfield  {journal}
  {\bibinfo  {journal} {JHEP}\ }\textbf {\bibinfo {volume} {04}},\ \bibinfo
  {pages} {056} (\bibinfo {year} {2016})},\ \Eprint
  {http://arxiv.org/abs/1601.01326} {arXiv:1601.01326 [hep-ph]} \BibitemShut
  {NoStop}%
\bibitem [{\citenamefont {Aad}\ \emph {et~al.}(2013)\citenamefont {Aad} \emph
  {et~al.}}]{Aad:2013yna}%
  \BibitemOpen
  \bibfield  {author} {\bibinfo {author} {\bibfnamefont {G.}~\bibnamefont
  {Aad}} \emph {et~al.} (\bibinfo {collaboration} {ATLAS}),\ }\href {\doibase
  10.1103/PhysRevD.88.112006} {\bibfield  {journal} {\bibinfo  {journal} {Phys.
  Rev.}\ }\textbf {\bibinfo {volume} {D88}},\ \bibinfo {pages} {112006}
  (\bibinfo {year} {2013})},\ \Eprint {http://arxiv.org/abs/1310.3675}
  {arXiv:1310.3675 [hep-ex]} \BibitemShut {NoStop}%
\bibitem [{\citenamefont {Khachatryan}\ \emph {et~al.}(2015)\citenamefont
  {Khachatryan} \emph {et~al.}}]{CMS:2014gxa}%
  \BibitemOpen
  \bibfield  {author} {\bibinfo {author} {\bibfnamefont {V.}~\bibnamefont
  {Khachatryan}} \emph {et~al.} (\bibinfo {collaboration} {CMS}),\ }\href
  {\doibase 10.1007/JHEP01(2015)096} {\bibfield  {journal} {\bibinfo  {journal}
  {JHEP}\ }\textbf {\bibinfo {volume} {01}},\ \bibinfo {pages} {096} (\bibinfo
  {year} {2015})},\ \Eprint {http://arxiv.org/abs/1411.6006} {arXiv:1411.6006
  [hep-ex]} \BibitemShut {NoStop}%
\bibitem [{\citenamefont {Shirai}\ \emph {et~al.}(2009)\citenamefont {Shirai},
  \citenamefont {Takahashi},\ and\ \citenamefont {Yanagida}}]{Shirai:2009fq}%
  \BibitemOpen
  \bibfield  {author} {\bibinfo {author} {\bibfnamefont {S.}~\bibnamefont
  {Shirai}}, \bibinfo {author} {\bibfnamefont {F.}~\bibnamefont {Takahashi}}, \
  and\ \bibinfo {author} {\bibfnamefont {T.~T.}\ \bibnamefont {Yanagida}},\
  }\href {\doibase 10.1016/j.physletb.2009.09.049} {\bibfield  {journal}
  {\bibinfo  {journal} {Phys. Lett.}\ }\textbf {\bibinfo {volume} {B680}},\
  \bibinfo {pages} {485} (\bibinfo {year} {2009})},\ \Eprint
  {http://arxiv.org/abs/0905.0388} {arXiv:0905.0388 [hep-ph]} \BibitemShut
  {NoStop}%
\bibitem [{\citenamefont {Bhattacherjee}\ \emph {et~al.}(2013)\citenamefont
  {Bhattacherjee}, \citenamefont {Evans}, \citenamefont {Ibe}, \citenamefont
  {Matsumoto},\ and\ \citenamefont {Yanagida}}]{Bhattacherjee:2013gr}%
  \BibitemOpen
  \bibfield  {author} {\bibinfo {author} {\bibfnamefont {B.}~\bibnamefont
  {Bhattacherjee}}, \bibinfo {author} {\bibfnamefont {J.~L.}\ \bibnamefont
  {Evans}}, \bibinfo {author} {\bibfnamefont {M.}~\bibnamefont {Ibe}}, \bibinfo
  {author} {\bibfnamefont {S.}~\bibnamefont {Matsumoto}}, \ and\ \bibinfo
  {author} {\bibfnamefont {T.~T.}\ \bibnamefont {Yanagida}},\ }\href {\doibase
  10.1103/PhysRevD.87.115002} {\bibfield  {journal} {\bibinfo  {journal} {Phys.
  Rev.}\ }\textbf {\bibinfo {volume} {D87}},\ \bibinfo {pages} {115002}
  (\bibinfo {year} {2013})},\ \Eprint {http://arxiv.org/abs/1301.2336}
  {arXiv:1301.2336 [hep-ph]} \BibitemShut {NoStop}%
\bibitem [{\citenamefont {Bomark}\ \emph {et~al.}(2009)\citenamefont {Bomark},
  \citenamefont {Lola}, \citenamefont {Osland},\ and\ \citenamefont
  {Raklev}}]{Lola:2008bk}%
  \BibitemOpen
  \bibfield  {author} {\bibinfo {author} {\bibfnamefont {N.~E.}\ \bibnamefont
  {Bomark}}, \bibinfo {author} {\bibfnamefont {S.}~\bibnamefont {Lola}},
  \bibinfo {author} {\bibfnamefont {P.}~\bibnamefont {Osland}}, \ and\ \bibinfo
  {author} {\bibfnamefont {A.~R.}\ \bibnamefont {Raklev}},\ }\href {\doibase
  10.1016/j.physletb.2009.05.011} {\bibfield  {journal} {\bibinfo  {journal}
  {Phys. Lett.}\ }\textbf {\bibinfo {volume} {B677}},\ \bibinfo {pages} {62}
  (\bibinfo {year} {2009})},\ \Eprint {http://arxiv.org/abs/0811.2969}
  {arXiv:0811.2969 [hep-ph]} \BibitemShut {NoStop}%
\bibitem [{\citenamefont {Bomark}\ \emph {et~al.}(2010)\citenamefont {Bomark},
  \citenamefont {Lola}, \citenamefont {Osland},\ and\ \citenamefont
  {Raklev}}]{Bomark:2009zm}%
  \BibitemOpen
  \bibfield  {author} {\bibinfo {author} {\bibfnamefont {N.~E.}\ \bibnamefont
  {Bomark}}, \bibinfo {author} {\bibfnamefont {S.}~\bibnamefont {Lola}},
  \bibinfo {author} {\bibfnamefont {P.}~\bibnamefont {Osland}}, \ and\ \bibinfo
  {author} {\bibfnamefont {A.~R.}\ \bibnamefont {Raklev}},\ }\href {\doibase
  10.1016/j.physletb.2010.02.050} {\bibfield  {journal} {\bibinfo  {journal}
  {Phys. Lett.}\ }\textbf {\bibinfo {volume} {B686}},\ \bibinfo {pages} {152}
  (\bibinfo {year} {2010})},\ \Eprint {http://arxiv.org/abs/0911.3376}
  {arXiv:0911.3376 [hep-ph]} \BibitemShut {NoStop}%
\bibitem [{\citenamefont {Affleck}\ \emph {et~al.}(1984)\citenamefont
  {Affleck}, \citenamefont {Dine},\ and\ \citenamefont
  {Seiberg}}]{Affleck:1983mk}%
  \BibitemOpen
  \bibfield  {author} {\bibinfo {author} {\bibfnamefont {I.}~\bibnamefont
  {Affleck}}, \bibinfo {author} {\bibfnamefont {M.}~\bibnamefont {Dine}}, \
  and\ \bibinfo {author} {\bibfnamefont {N.}~\bibnamefont {Seiberg}},\ }\href
  {\doibase 10.1016/0550-3213(84)90058-0} {\bibfield  {journal} {\bibinfo
  {journal} {Nucl. Phys.}\ }\textbf {\bibinfo {volume} {B241}},\ \bibinfo
  {pages} {493} (\bibinfo {year} {1984})}\BibitemShut {NoStop}%
\end{thebibliography}%

\end{document}